# Modeling of electromagnetic wave propagation and spectra of optical excitations in complex media using $4 \times 4$ matrix formalism

P. D. Rogers,[1,*] M. Kotelyanskii,[1,2] and A. A. Sirenko[1]

[1] *Department of Physics, New Jersey Institute of Technology, Newark, New Jersey 07102, USA*

[2] *Rudolph Technologies Inc., Flanders, New Jersey 07836, USA*

[*] pdr2@njit.edu

Using $4 \times 4$ matrix formalism we analyzed electromagnetic wave propagation and Jones matrix components for reflectivity and transmittivity in bi-anisotropic materials. Analytic formulas for complex reflection and transmission coefficients for bi-anisotropic materials in both semi-infinite and thin-film configurations have been derived. The obtained results are applicable for analysis of the optical spectra of multiferroic crystals and metamaterials. The Adjusted Oscillator Strength Matching Condition (AOSM) for hybrid magnetic- and electric-dipole excitations in anisotropic multiferroics is derived for oblique angles of incidence. Mueller Matrices are used to simulate spectra of magneto-electric and chiral excitations and methods to distinguish them are discussed.

# 1. INTRODUCTION

## A. Motivation

Optical spectra of complex materials, such as magneto-electric (ME) and multiferroic crystals, materials with intrinsic or artificial chirality, and metamaterials, are in the focus of modern experimental and theoretical studies. The common feature of these complex materials is that their optical properties cannot be described only with a 3×3 dielectric susceptibility tensor $\hat{\varepsilon}(\omega)$. The complex materials can reveal a so-called *bi-anisotropic*[1] optical behavior in a form of the fascinating effects, such as nonreciprocal light propagation, negative index of refraction (NIR), and polarization plane rotation. These exotic optical phenomena usually occur in a relatively narrow part of the optical spectrum. For example, the NIR effect could occur in metamaterials or multiferroics only in the GHz or THz spectral ranges, but above a certain frequency such materials behave as normal metals or dielectrics and, hence, a simple $\varepsilon(\omega)$ function could perfectly describe their optical properties in, for example, the visible part of the spectrum. Note also that the bi-anisotropic optical phenomena, such as magneto-electric and chirality effects, are not mutually exclusive and can coexist in the same or different parts of the optical spectrum. The proper description of the bi-anisotropic optical effects in complex materials requires an adequate theoretical description and advanced experimental spectroscopic approaches. Calculations of the optical *spectra* and *polarization* for complex anisotropic materials are particularly important at the resonance with the related elementary excitations, such as electromagnons in magneto-electric materials.

Recently, spectra of electromagnons in $TbMnO_3$ and $GdMnO_3$ multiferroic crystals have been discovered by Pimenov *et al.*[2] Similar electromagnon excitations have been also observed in other related multiferroic oxides, including $RE MnO_3$ and $RE Mn_2O_5$ ($RE$ = rare earth elements).[3-6] The polarization selection-rules analysis for the transmission optical configuration suggested that the electromagnon mode is excited by an electric field of light, in contrast to the case of antiferromagnetic resonances (AFMR) that can be excited by the magnetic field component of light only. However, the polarization analysis of the electromagnon spectra has been always restricted by the experimental geometry with the normal light incidence on the sample surface. The limitations of this approach revealed themselves recently by failing to explain the experimentally-observed suppression of electromagnons in reflectivity measurements of $GdMnO_3$.[7] As we will see in the following, electromagnons in uniaxial crystals are not optically active in a back-reflection configuration, while transmission technique applied alone is not capable to differentiate between pure magnetic- and electric dipoles and, of course, cannot distinguish them from electromagnon-type of excitations. As we demonstrate in this paper, the theoretical representation of bi-anisotropic phenomena can be done with the help of the Jones and/or Mueller Matrices (MM). Correspondingly, the most efficient experimental technique for the experimental studies of complex materials is Mueller Matrix Spectroscopic Ellipsometry (MMSE) that can be realized in both reflection and transmission configurations with variable azimuthal angle and variable angle of incidence (AOI).

We apply $4 \times 4$ Berreman's matrix formalism to calculate polarization of the optical spectra in complex materials. Analytic formulas for reflection and transmission coefficients for bi-anisotropic materials in both semi-infinite and thin-film configurations have been derived using *Mathematica* symbolic language and verified against the exact numeric code. We illustrate our results with examples of Mueller Matrix calculations such as the spectra of electromagnons in multiferroic materials in the far-infrared spectral range. These calculations are also applicable to a diverse group of bi-anisotropic materials, such as metamaterials and chiral structures with different types of atomic interactions responsible for their non-trivial optical properties. Our results should provide a foundation for building the adequate forward models that can be used in the experimental data analysis obtained with MMSE and other spectroscopic techniques, such as Rotating Analyzer Ellipsometry, Generalized Ellipsometry, and Transmission Polarimetry.

## B. Modeling Approach

Models for electromagnetic wave propagation in a medium require solutions to Maxwell's equations. These solutions, in turn, depend upon the proper characterization of the electromagnetic properties of the medium. As Weiglhofer explains in a theatrical analogy, if Maxwell's equations are a play with intricate plots, then the medium is the stage in which the electromagnetic processes take place.[1] The stage is described by a set of equations which are known as the medium's constitutive relations:

$$\vec{D} = \hat{\varepsilon}\vec{E} + \hat{\rho}\vec{H}$$
$$\vec{B} = \hat{\rho}'\vec{E} + \hat{\mu}\vec{H}$$

(1)

In Eq. (1), $\vec{D}$ is the dielectric displacement, $\vec{B}$ is the magnetic induction, $\vec{E}$ is the primary electric field vector, $\vec{H}$ is the primary magnetic field vector, $\hat{\varepsilon}$ is the dielectric permittivity tensor, $\hat{\mu}$ is the magnetic permeability tensor, and $\hat{\rho}$ and $\hat{\rho}'$ are the bi-anisotropic tensors.[8] Each tensor is associated with a unique physical property of the medium and can be described by a $3 \times 3$ matrix.[9] Further, the physical properties of the medium, which the tensors represent, are often frequency-dependent and must be described by a set of dispersion equations. Various models including the simple harmonic oscillator (SHO) and coupled harmonic oscillator (CHO) models are usually used for these dispersion relationships. Some of these models will be considered below. In this paper, a *simple medium* is defined to have isotropic $\hat{\varepsilon}$ and $\hat{\mu}$ tensors and no bi-anisotropic activity. A *complex medium* will refer to all other possible tensor symmetries and allowed tensor combinations.[1] In this paper we do not consider the effects of non-linearity nor spatial dispersion. The combination of Maxwell's equations, boundary conditions, the constitutive relations, and the dispersion relations are required to derive a proper solution for electromagnetic wave propagation and to model excitations in the optical spectra.

In contrast to $\hat{\varepsilon}$ and $\hat{\mu}$ tensors, the bi-anisotropic tensors $\hat{\rho}$ and $\hat{\rho}'$ are less known and their properties require clarification. In this paper we will consider two major additive contributions to $\hat{\rho}$ and $\hat{\rho}'$: the magneto-electric effect and chirality, so that:

$$\hat{\rho} = \hat{\alpha} + i \cdot \hat{\xi}$$
$$\hat{\rho}' = \hat{\alpha}' - i \cdot \hat{\xi}^T \qquad (2)$$

One can see that the ME contribution is described by the complex tensor $\hat{\alpha}$, and chirality is represented by the tensor $\hat{\xi}$. Both tensors, $\hat{\xi}$ and $\hat{\alpha}$, are complex and can have both real and imaginary parts. Also, $\hat{\rho}$ and $\hat{\rho}'$ are not expected to be the complex-conjugate-transpose for each other.[8]

According to Dzyaloshinskii, based on the thermodynamic stability argument, the corresponding ME contribution to $\hat{\rho}'$ should be a symmetric tensor: $\hat{\alpha}' = \hat{\alpha}^T$ in the static case when $\omega \to 0$. This requirement of $\hat{\alpha}' = \hat{\alpha}^T$ for optical spectra is under debate in the literature.[10] In the following theoretical analysis, we keep a general notation for the $\hat{\rho}$ and $\hat{\rho}'$ tensors. In any case, both $\hat{\alpha}$ and $\hat{\alpha}'$ have the same sign of their complex parts, so that the material should absorb radiation, and not generate it. Tensors $\hat{\alpha}$ and $\hat{\alpha}'$ change sign under space inversion and time inversion operation, remaining unchanged if both operations are applied simultaneously.

This property results in the requirement that $\hat{\alpha} = \hat{\alpha}' \equiv 0$ in materials with the center of inversion or with time-reversal symmetry (see Ref. [11, 12] for more detail). In contrast to $\hat{\alpha}$, the chirality term $i \cdot \hat{\xi}$ has its transpose-complex conjugate counterpart $-i \cdot \hat{\xi}^T$ that contributes to $\hat{\rho}'$. For isotropic materials, Georgieva[13] showed that the chirality parameter $\xi$, which originates from the $\partial \vec{H} / \partial t$ and $\partial \vec{E} / \partial t$ terms in the Maxwell equations, scales proportionally to $\omega$ and disappears at zero frequency: $\xi(0) \to 0$. In the following we are not imposing any restrictions on $\xi(\omega)$ and one could assume a resonant behavior for a complex function $\xi(\omega)$ and diminishing of chirality effects at high frequencies: $\xi(\infty) \to 0$.

The task of obtaining analytical solutions for all possible configurations appears daunting, but $4 \times 4$ matrix formalism, as developed by Berreman,[14] provides for an accurate and systematic method of obtaining numerical, and in some cases, analytic solutions for electromagnetic wave propagation in both simple and complex media. A complete description of electromagnetic wave propagation in a complex medium is made possible using Berreman's matrix equation: [14]

$$\begin{pmatrix} 0 & -curl \\ curl & 0 \end{pmatrix} \begin{pmatrix} \vec{E} \\ \vec{H} \end{pmatrix} = i \frac{\omega}{c} \begin{pmatrix} \hat{\varepsilon} & \hat{\rho} \\ \hat{\rho}' & \hat{\mu} \end{pmatrix} \begin{pmatrix} \vec{E} \\ \vec{H} \end{pmatrix} \tag{3}$$

In Eq. (3), *curl* represents the 3×3 matrix operator. The first matrix on the right hand side is a 6×6 matrix called the optical matrix **M**. This matrix contains all information about the constitutive relations and completely describes the anisotropic properties of the material.[15] Eq. (3) can be reduced to the Berreman equation which describes electromagnetic wave propagation in a crystal:

$$\frac{d\Psi}{dz} = i \frac{\omega}{c} \tilde{\Delta} \Psi \tag{4}$$

In Eq. (4), $\Psi$ is an array of the transverse components of the electromagnetic wave $[E_x, H_y, E_y, -H_x]^T$ in the medium and is an eigenvector of $\tilde{\Delta}$, where $\tilde{\Delta}$ is a $4 \times 4$ matrix constructed from the components of the $\hat{\varepsilon}$, $\hat{\mu}$, $\hat{\rho}$ and $\hat{\rho}'$ tensors. Eq. (4) is at the heart of $4 \times 4$ matrix formalism. The eigenvalue and eigenvector solutions to Eq. (4) represent wave vectors and the transverse components of the propagating electromagnetic waves, respectively. These solutions are unique to the crystal symmetries and constitutive relations incorporated into the $\tilde{\Delta}$ matrix.

Various combinations of the tensors contained in Eq. (1), Eq. (2) and Eq. (3) and their allowed symmetries have been studied in the literature. We organize this topic beginning with papers addressing the $\hat{\varepsilon}$ tensor and then cite additional references as the tensors $\hat{\mu}$, $\hat{\alpha}$ and $\hat{\xi}$ are added. Schubert[16, 17] and Schubert et al.[18] studied anisotropic $\hat{\varepsilon}$ using Generalized Ellipsometry and applied this work to layered structures. Wohler et al.[19] used $4 \times 4$ matrix formalism to calculate the transfer matrix for a material with $\hat{\varepsilon}$ having uniaxial symmetry. Mayerhofer et al.[20] employed $4 \times 4$ matrix formalism using anisotropic $\hat{\varepsilon}$ tensors to calculate the reflection coefficients for semi-infinite crystals with monoclinic symmetry. Veselago[21] and Veselago et al.[22] addressed the case of isotropic $\hat{\varepsilon}$ and $\hat{\mu}$ for the semi-infinite configuration and considered the important phenomenon of NIR. Grzegorczyk et al.[23] studied the refraction laws for metamaterials having anisotropic $\hat{\varepsilon}$ and $\hat{\mu}$ in the semi-infinite configuration. Smith et al.[24-26] considered anisotropic $\hat{\varepsilon}$ and $\hat{\mu}$ for both semi-infinite indefinite media and layered metamaterial structures while Driscoll et al [27, 28] analyzed the complex reflection and transmission coefficients for metamaterials with anisotropic $\hat{\varepsilon}$ and $\hat{\mu}$ in the thin film configuration. Combinations of $\hat{\varepsilon}$, $\hat{\mu}$ and the chirality tensor $\hat{\xi}$ have also been studied extensively in the literature. Arteaga et al.

[29-33] and Bahar et al.[34-38] have employed Mueller Matrices in their analysis of chirality. Bassiri et al. [39] and later Cory et al.[40] calculate the complex reflection coefficients for a chiral medium with isotropic $\hat{\varepsilon}$, $\hat{\mu}$ and $\hat{\xi}$. Georgieva[13] and Georgieva et al.[41] employed both $4 \times 4$ matrix formalism and Mueller Matrices in their analysis of reflection from chiral media with isotropic $\hat{\varepsilon}$ and $\hat{\xi}$. Konstantinova et al.[42, 43] used $4 \times 4$ matrix formalism to analyze a number of crystal characteristics including optical activity for anisotropic $\hat{\varepsilon}$, $\hat{\mu}$ and $\hat{\xi}$. Cheng and Cui[44, 45] calculated the wave vectors and Poynting vectors for a chiral medium with uniaxial $\hat{\varepsilon}$ and $\hat{\mu}$. The allowed symmetries for anisotropic $\hat{\varepsilon}$, $\hat{\mu}$ and the magneto-electric tensor $\hat{\alpha}$ have been addressed by O'Dell[11] and Rivera[12], as well as being extended to the discussion of electromagnons in rare earth manganites by Cano[10]. The case of bi-isotropic materials consisting of isotropic $\hat{\varepsilon}$, $\hat{\mu}$, $\hat{\alpha}$ and $\hat{\xi}$ is discussed extensively in Lindell et al.[8]. Lindell et al.[8] also calculate the wave vectors for a bi-anisotropic medium where each of the four tensors have uniaxial symmetry. This paper adds to this literature by demonstrating that, using $4 \times 4$ matrix formalism and Mueller Matrices, the complex reflection and transmission coefficients and their associated reflection and transmission intensities can be calculated and simulated for an arbitrary bi-anisotropic medium in both the semi-infinite and thin film configuration.

The main challenge to the analysis of bi-anisotropic materials is a vast number of possible tensor symmetries in the bulk crystals and thin films. In this paper, five different configurations of crystal symmetry and constitutive tensors will be examined. They are presented in increasing order of complexity. The five cases have been selected to illustrate both the application of $4 \times 4$ matrix formalism as well as aspects of electromagnetic wave propagation.

Case 1 examines a medium with anisotropic $\hat{\varepsilon}$ and $\hat{\mu}$ tensors only $(\hat{\rho} = \hat{\rho}' = 0)$. This case is applicable, for example, to a system with a center of inversion or with time-reverse invariance. We consider this as the base case of our analysis because it illustrates how the eigenvalues of the $\tilde{\Delta}$ matrix are evident not only in the eigenvectors describing the electromagnetic waves but also in the complex reflection coefficients, $\vec{k}$ vectors and Poynting vectors associated with each polarization. Case 1 illustrates how material anisotropy reveals itself in the non-degenerate eigenvalue solutions of the $\tilde{\Delta}$ matrix.

Case 2 examines isotropic $\hat{\varepsilon}$ and $\hat{\mu}$ tensors. It is an immediate consequence of Case 1. We compare Case 2 results obtained using $4 \times 4$ matrix formalism to the Veselago approach for materials having non-trivial magnetic permeability $\mu \neq 1$.[21]

Case 3 introduces magneto-electric tensors into the analysis by examining the case of a material with uniaxial $\hat{\varepsilon}$ and $\hat{\mu}$ tensors and magneto-electric tensors having only one off-diagonal element. This is typical for multiferroic materials. This scenario also presents the interesting case of irreversibility in electro-magnetic wave propagation in magneto-electric crystals.[11]

In Case 4, the analysis of isotropic $\hat{\varepsilon}$, $\hat{\mu}$, $\hat{\rho}$ and $\hat{\rho}'$ tensors is presented. Solutions for this symmetry are more mathematically complicated compared to the first three cases yet still permit analytic solutions. Case 4 illustrates how birefringence typical for anisotropic materials can be introduced into a material with isotropic $\hat{\varepsilon}$ and $\hat{\mu}$ tensors through the magneto-electric effect.

Finally, Case 5 analyzes anisotropic $\hat{\varepsilon}$, $\hat{\mu}$, $\hat{\rho}$ and $\hat{\rho}'$ tensors, all in orthorhombic symmetry. Case 5 illustrates how the $\tilde{\Delta}$ matrix can be constructed for such complicated constitutive relations. Case 5 will be illustrated using results of our numerical analysis. A simulation tool that covers reflectivity geometry for semi-infinite bi-anisotropic material is available in Ref. [46].

In Section II, each of the five cases is analyzed for the semi-infinite configuration. The analysis follows the flowchart for $4 \times 4$ matrix formalism outlined in Fig. 1. This procedure begins with the **M** matrix which enables the $\tilde{\Delta}$ matrix to be calculated along with its eigenvalues and eigenvectors. From the eigenvalues, the $\vec{k}$ vectors can be immediately determined which, in turn, allow for the analysis of possible birefringence in the medium. The eigenvectors, together with the tangential boundary conditions on $\vec{E}$ and $\vec{H}$ for non-magnetic incident media, provide for the solution of the complex reflection coefficients. Finally, when the $z$ components of $\vec{E}$ and $\vec{H}$ are recovered, the Poynting vector is returned, which can then be compared to the wave vector for analysis of possible divergence between the direction of the wave fronts and energy flow.

In Section III, the procedure is applied to the thin film configuration for Cases 1 and 3, where the method takes into account interference from the multiply reflected waves at the surface boundaries. The analysis of bi-anisotropic materials in thin film configuration also allows for the calculation of the complex transmission coefficients assuming both a non-magnetic incident medium (usually but not necessarily vacuum) and a non-magnetic substrate. In Section IV, the need for dispersion models for proper modeling of the response optical functions is examined. For Case 3, which incorporates the magneto-electric effect, the implications of dispersion for the optical wave propagation and NIR are discussed. In Section V, the interesting case of hybrid modes, *i.e.*, electric and magnetic excitations at the same resonant frequency, is examined for variable angle of light incidence. Using the complex reflection and transmission formulas derived in previous Sections, together with the dispersion models, the condition called the Adjusted Oscillator Strength Matching is discussed for anisotropic media. Under this condition, we show that the hybrid modes can disappear in the Reflectivity spectra but still remain strong in the Transmission spectra.[47] Finally, in Section VI, we simulate electric, magnetic, hybrid and electromagnon modes in the Reflectivity spectra. Mueller Matrices are used to illustrate these simulations in both the frequency and AOI domains. Full MM analysis allows for the possibility of distinguishing between many of the electric, magnetic and magneto-electric effects.

## II. SEMI-INFINITE CONFIGURATION

### A. Case 1 - Orthorhombic $\hat{\varepsilon}$ and $\hat{\mu}$ Tensors; ($\rho = \rho' = 0$)

The case of a material having orthorhombic $\hat{\varepsilon}$ and $\hat{\mu}$ tensors will now be examined. It is assumed that this crystal has principal axes parallel to the $x, y$ and $z$ coordinate axes which form a right hand system with the $z$ axis pointing downwards and the $x$ axis pointing to the right. Radiation is incident in the $x - z$ plane. This configuration is illustrated in Fig. 2. [48]

We further assume that the $\hat{\varepsilon}$ and $\hat{\mu}$ tensors can be simultaneously diagonalized in the same $x$ - $y$ - $z$ coordinate system. In crystals, the principal axes for $\hat{\varepsilon}$ and $\hat{\mu}$ tensors rarely coincide. Accordingly, this symmetry realization is mostly applicable to metamaterials. With no magneto-electric activity, the optical matrix **M** becomes:

$$\begin{pmatrix} \varepsilon_{xx} & 0 & 0 & 0 & 0 & 0 \\ 0 & \varepsilon_{yy} & 0 & 0 & 0 & 0 \\ 0 & 0 & \varepsilon_{zz} & 0 & 0 & 0 \\ 0 & 0 & 0 & \mu_{xx} & 0 & 0 \\ 0 & 0 & 0 & 0 & \mu_{yy} & 0 \\ 0 & 0 & 0 & 0 & 0 & \mu_{zz} \end{pmatrix}, \qquad (5)$$

and $\tilde{\Delta}$ is a $4 \times 4$ matrix calculated to be:[14]

$$\tilde{\Delta} = \begin{pmatrix} 0 & \mu_{yy} - \dfrac{N_0{}^2 \sin(\theta_0)^2}{\varepsilon_{zz}} & 0 & 0 \\ \varepsilon_{xx} & 0 & 0 & 0 \\ 0 & 0 & 0 & \mu_{xx} \\ 0 & 0 & \varepsilon_{yy} - \dfrac{N_0{}^2 \sin(\theta_0)^2}{\mu_{zz}} & 0 \end{pmatrix} (6)$$

Inserting Eq. (5) and Eq. (6) into Eq. (3) returns four exact solutions of the form $\psi_l(z) = \psi_l(0) e^{i q_l z}$ with $l = 1, 2, 3$ or $4$, two for each of the $p$ and $s$ polarization states. $\theta_0$ is the angle of incidence (AOI), while $p(s)$ refers to radiation parallel (perpendicular) to the plane of incidence. $q_{zp}$ and $q_{zs}$ are the eigenvalues associated with $p$ and $s$ polarizations, respectively and constitute the $z$ components of the wave vectors in the medium as

$$q_{zp} = \pm \frac{\omega}{c} \sqrt{\varepsilon_{xx}} \sqrt{\mu_{yy} - \frac{N_0{}^2 \sin^2(\theta_0)}{\varepsilon_{zz}}}$$

$$q_{zs} = \pm \frac{\omega}{c} \sqrt{\mu_{xx}} \sqrt{\varepsilon_{yy} - \frac{N_0{}^2 \sin^2(\theta_0)}{\mu_{zz}}}$$

(7)

Given the $z$ components of the wave vector in Eq. (7), the complete wave vectors for each of the

$p$ and $s$ polarization states can be written as:

$$\vec{k}_p = \left( \left(\frac{\omega}{c}\right) N_0 \sin(\theta_0), 0, \frac{\omega}{c} \sqrt{\varepsilon_{xx}} \sqrt{\mu_{yy} - \frac{N_0{}^2 \sin^2(\theta_0)}{\varepsilon_{zz}}} \right)$$

$$\vec{k}_s = \left( \left(\frac{\omega}{c}\right) N_0 \sin(\theta_0), 0, \frac{\omega}{c} \sqrt{\mu_{xx}} \sqrt{\varepsilon_{yy} - \frac{N_0{}^2 \sin^2(\theta_0)}{\mu_{zz}}} \right)$$

(8)

The two $k$ vectors in Eq. (8) identify the direction of propagation of the waves associated with each polarization. It is clear that for non zero AOI, the two $k$ vectors will not be identical and will therefore diverge as they propagate forward (downward) into the medium. This phenomenon is known as birefringence and is evidenced by two separate forward propagating electromagnetic waves.

The eigenvector solutions (in columns) are:

$$
\begin{pmatrix}
1 & 0 & 1 & 0 \\[4pt]
\dfrac{\sqrt{\varepsilon_{xx}}}{\sqrt{\mu_{yy}-\dfrac{N_0^2\sin\left(\theta_0\right)^2}{\varepsilon_{zz}}}} & 0 & -\dfrac{\sqrt{\varepsilon_{xx}}}{\sqrt{\mu_{yy}-\dfrac{N_0^2\sin\left(\theta_0\right)^2}{\varepsilon_{zz}}}} & 0 \\[10pt]
0 & 1 & 0 & 1 \\[4pt]
0 & \dfrac{\sqrt{\varepsilon_{yy}-\dfrac{N_0^2\sin\left(\theta_0\right)^2}{\mu_{zz}}}}{\sqrt{\mu_{xx}}} & 0 & -\dfrac{\sqrt{\varepsilon_{yy}-\dfrac{N_0^2\sin\left(\theta_0\right)^2}{\mu_{zz}}}}{\sqrt{\mu_{xx}}}
\end{pmatrix}
$$

$$(9)$$

In Eq. (9), the eigenvectors in columns 1 and 2 represent forward propagating waves while those in columns three and four represent backward propagating waves. The eigenvectors in columns one and three are associated with the $q_{zp}$ eigenvalue and represent $p$ polarized radiation. A complete description of this wave involves multiplication by $\exp\left(\pm i q_{zp} z\right)$. Similarly, the eigenvectors in columns two and four are associated with the $q_{zs}$ eigenvalue and represent $s$ polarized radiation. A complete description of this wave involves multiplication by $\exp\left(\pm i q_{zs} z\right)$. For a semi-infinite material, the two eigenvectors representing the forward propagating waves are used to calculate the complex reflection coefficients for $p$ and $s$ polarized radiation. The procedure for calculating the complex reflection coefficients involves matching the tangential components of the incident and reflected $\vec{E}$ and $\vec{H}$ fields to a linear combination of the two eigenvectors calculated at the common interface located at $z = 0$.[14, 15] The complex reflection coefficients are:

$$r_{pp} = \frac{\varepsilon_{xx}k_{z0} - N_0{}^2 q_{zp}}{\varepsilon_{xx}k_{z0} + N_0{}^2 q_{zp}} \qquad (10)$$

$$r_{ss} = \frac{\mu_{xx} k_{z0} - q_{zs}}{\mu_{xx} k_{z0} + q_{zs}}. \qquad (11)$$

In Eq. (10) and Eq. (11), $k_{z0} = \frac{\omega}{c} N_0 \cos(\theta_0)$ is the $z$ component of the incident wave vector and $N_0$ is the index of refraction in the incident medium. The eigenvectors in Eq. (9) can also be used to calculate the Poynting vector for each of the $p$ and $s$ polarized radiation states. This procedure first requires recapture of the $z$ components of the $\vec{E}$ and $\vec{H}$ fields which were originally suppressed in the Berreman equations in order to reduce from a $6 \times 6$ to a $4 \times 4$ formalism. By solving the two algebraic equations associated with the initial Berreman matrices, for orthorhombic symmetry, the solutions for the $z$ components are:

$$E_z = -\frac{H_y N_0 \sin(\theta_0)}{\varepsilon_{zz}}$$

$$(12)$$

$$H_z = \frac{E_y N_0 \sin(\theta_0)}{\mu_{zz}}$$

Eq. (12) can be applied to each of the $p$ and $s$ polarization states. Since the terms in Eq. (9) recur frequently in this analysis, we define $\varsigma = \sqrt{\varepsilon_{yy} - \frac{N_0^2 \sin(\theta_0)^2}{\mu_{zz}}}$ and $\eta = \sqrt{\mu_{yy} - \frac{N_0^2 \sin(\theta_0)^2}{\varepsilon_{zz}}}$.

First consider $p$ polarization. Here, $H_z$ becomes zero and the vector fields are:

$$\vec{E} = E_x \left(1, 0, \frac{-\sqrt{\varepsilon_{xx}} N_0 \sin(\theta_0)}{\varepsilon_{zz} \eta}\right) e^{iq_{zp}z}$$

$$(13)$$

$$\vec{H} = E_x \left(0, \frac{\sqrt{\varepsilon_{xx}}}{\eta}, 0\right) e^{iq_{zp}z}$$

The fields in Eq. (13) now permit the calculation of the Poynting vector $\vec{S} = \frac{1}{2}\left(\vec{E} \times \vec{H}^{*}\right)$ applicable to $p$ polarization:

$$\vec{S}_{p} = \frac{1}{2}\left|E_{x}\right|^{2}\left(\left|\frac{\sqrt{\varepsilon_{xx}}}{\eta}\right|^{2}\frac{N_{0}\sin\left(\theta_{0}\right)}{\varepsilon_{zz}}, 0, \left(\frac{\sqrt{\varepsilon_{xx}}}{\eta}\right)^{*}\right) \tag{14}$$

where the asterisks, *, represents the complex conjugate operation. From Eq. (14), the tangent of the Poynting vector angle in the medium is:

$$\tan\left(\theta_{\vec{S}}\right)_{p} = \frac{\sqrt{\varepsilon_{xx}}N_{0}\sin\left(\theta_{0}\right)}{\varepsilon_{zz}\eta} \tag{15}$$

From Eq. (8), the tangent of the $k$ vector angle in the medium is:

$$\tan\left(\theta_{\vec{k}}\right)_{p} = \frac{N_{0}\sin\left(\theta_{0}\right)}{\sqrt{\varepsilon_{xx}}\eta} \tag{16}$$

While the expressions in Eq. (15) and Eq. (16) are similar, a comparison shows that if $\varepsilon_{xx} \neq \varepsilon_{zz}$ they are not identical. This analysis points out the well known observation that for a crystal with an orthorhombic symmetry, the direction of the wave vector is not identical to that of the energy flow as given by the Poynting vector. For $s$ polarization, $E_{z}$ is zero and the fields become:

$$\vec{E} = E_y \left(0, 1, 0\right) e^{iq_{zs}z}$$

$$\vec{H} = E_y \left(-\frac{\varsigma}{\sqrt{\mu_{xx}}}, 0, \frac{N_0 \sin\left(\theta_0\right)}{\mu_{zz}}\right) e^{iq_{zs}z}$$

(17)

and the Poynting vector for $s$ polarization is found to be:

$$\vec{S} = \frac{1}{2} \left|E_y\right|^2 \left(\left(\frac{N_0 \sin\left(\theta_0\right)}{\mu_{zz}}\right)^*, 0, \left(\frac{\varsigma}{\sqrt{\mu_{xx}}}\right)^*\right).$$

(18)

From Eq. (18), the tangent of the Poynting vector angle for $s$ polarization is:

$$\tan\left(\theta_{\vec{S}}\right)_s = \frac{\sqrt{\mu_{xx}} N_0 \sin\left(\theta_0\right)}{\mu_{zz}\varsigma},$$

(19)

and from Eq. (8), the tangent of the $k$ vector angle for $s$ polarization is calculated to be:

$$\tan\left(\theta_{\vec{k}}\right)_s = \frac{N_0 \sin\left(\theta_0\right)}{\sqrt{\mu_{xx}}\varsigma}.$$

(20)

Again, a comparison between Eq. (20) and Eq. (19) shows that if $\mu_{xx} \neq \mu_{zz}$ the two angles $\theta_{\vec{k}}$ and $\theta_{\vec{S}}$ are not identical. Accordingly, the $s$ polarized state will also experience a divergence between the direction of wave propagation in the crystal and the direction of energy flow. In summary, a crystal with orthorhombic $\hat{\varepsilon}$ and $\hat{\mu}$ tensors will give rise to four unique vectors: one unique $\vec{k}$ vector for each polarization and one unique $\vec{S}$ vector for each polarization, neither of which is coincident with its corresponding wave vector. These four vectors are simulated in Fig. 3(b) for an imaginary material with diagonal tensor components: $\varepsilon = (4,6,8)$ and $\mu = (1,2,3)$.

## B. Case 2-Isotropic $\varepsilon$ and $\mu$ Tensors ($\rho = \rho' = 0$)

Case 2 deals with a simple medium described by isotropic $\hat{\varepsilon}$ and $\hat{\mu}$ tensors. The **M** matrix for isotropic symmetry is given by:

$$\begin{pmatrix} \varepsilon & 0 & 0 & 0 & 0 & 0 \\ 0 & \varepsilon & 0 & 0 & 0 & 0 \\ 0 & 0 & \varepsilon & 0 & 0 & 0 \\ 0 & 0 & 0 & \mu & 0 & 0 \\ 0 & 0 & 0 & 0 & \mu & 0 \\ 0 & 0 & 0 & 0 & 0 & \mu \end{pmatrix} \qquad (21)$$

Conclusions regarding this symmetry are immediately available from the previous case by setting $\varepsilon_{xx} = \varepsilon_{yy} = \varepsilon_{zz} = \varepsilon$ and $\mu_{xx} = \mu_{yy} = \mu_{zz} = \mu$. A key result is the degeneracy of the eigenvalues:

$$q_{zp} = \frac{\omega}{c}\sqrt{\varepsilon\mu - N_0{}^2 \sin^2(\theta_0)}$$

$$q_{zs} = \frac{\omega}{c}\sqrt{\varepsilon\mu - N_0{}^2 \sin^2(\theta_0)} \qquad (22)$$

According to Eq. (22), for an isotropic crystal, there will be no birefringence as existed for the orthorhombic symmetry of Case 1. Both electromagnetic waves will, of course, follow identical paths. In addition, from Eqs. (15), (16), (19) and (20) it is clear that the direction of energy flow is also identical to the direction of wave propagation. This configuration is simulated in Fig. 3(a) for a material with diagonal tensor components: $\varepsilon = (4,4,4)$ and $\mu = (2,2,2)$, where all four vectors are coincident.

This case also illustrates how Veselago's approach for materials with $\mu \neq 1$ is automatically incorporated into the results using $4 \times 4$ matrix formalism via the solution of Maxwell's equations. For radiation normally incident from vacuum, the complex reflection coefficient for $s$ polarized radiation reduces to:

$$r_{ss} = \frac{\sqrt{\dfrac{\mu}{\varepsilon}} - 1}{\sqrt{\dfrac{\mu}{\varepsilon}} + 1} \qquad (23)$$

For a non-magnetic material, Fresnel's reflection coefficient is given by $r_{ss} = \dfrac{n_1 - n_2}{n_1 + n_2}$ where $n = \sqrt{\varepsilon}$ [49]. However, for a semi-infinite isotropic magnetic material, Veselago explained that $n$ should not be replaced by $\sqrt{\varepsilon \mu}$ but rather by $\sqrt{\varepsilon / \mu} = 1/z$, where $z$ is the wave impedance [21, 22]. The formula for the reflection coefficient then becomes $r_{ss} = \dfrac{z_2 - z_1}{z_2 + z_1}$. This expression is identical to Eq. (23) which is derived using $4 \times 4$ matrix formalism.

### C. Case 3-Anisotropic $\hat{\varepsilon}$ and $\hat{\mu}$ Tensors; Off-diagonal Magneto-Electric Tensors.

In Case 3, we introduce the magneto-electric effect in which a polarization vector $\vec{P}$ may be induced by the application of magnetic field $\vec{H}$, and a magnetization $\vec{M}$ may be induced from the application of electric field $\vec{E}$ [12]. There is much debate surrounding the theoretical explanation of these coupling mechanisms at the atomic level. Here we model the effect through the magneto-electric tensors, $\hat{\rho}$ and $\hat{\rho}'$, which couple the dielectric and magnetic response functions of a magneto-electric material. In the case of zero chirality $(\xi = 0)$, $\hat{\rho} = \hat{\alpha}$ and $\hat{\rho}' = \hat{\alpha}'$. As mentioned in Introduction, in the static case, $\alpha'$ is the transpose of $\alpha$. This requirement follows from the Dzyaloshinsky's definition of $\hat{\alpha}$ in the static case:

$$\alpha_{ij} = \frac{\partial^2 F}{\partial E_i \partial H_j} \qquad (24)$$

For the dynamic case, this relationship does not necessarily hold.[10] We note that other variables can be used to describe magneto-electric tensors in the ($\vec{E}$, $\vec{B}$) basis.[10, 12] In this paper we use the ($\vec{E}$, $\vec{H}$) basis as it is the most convenient for application of the 4×4 formalism procedures. Crystal symmetry plays a critical role in correctly defining the **M** matrix for magneto-electric and multiferroic materials. For example, the requirement that $\hat{\rho} = \hat{\alpha} \neq 0$ infers a simultaneous absence of both center of inversion and the time-reverse invariance. In symmetry terms, these constraints limit the number of possible magnetic point groups to 58 where the magneto-electric effect is possible.[11] Recent theoretical studies have included derivations of magneto-electric symmetries for spiral magnetic ordering in hexagonal manganites $R$MnO$_3$. It has been shown that the magneto-electric tensor $\rho$ for a cycloidal spin ordering, such as found in $R$MnO$_3$ compounds ($R$=rare earth), has only one non-zero element $\rho_{xy}$ and the $\hat{\varepsilon}$ and $\hat{\mu}$ tensors are uniaxial. [9, 10, 50]

Recall that $p$ polarization at normal incidence is defined with $\vec{E} \parallel \hat{x}$ and $\vec{H} \parallel \hat{y}$ for the incident radiation. At oblique angles of incidence for $p$ polarization, $\vec{E}$ will also have a $z$ component. For the $\hat{\varepsilon}$ tensor we use: $\varepsilon_{xx} = \varepsilon_{yy} = \varepsilon_\perp$ and $\varepsilon_{zz} = \varepsilon_\parallel$; for the $\hat{\mu}$ tensor we use $\mu_{xx} = \mu_{yy} = \mu_\perp$ and $\mu_{zz} = \mu_\parallel$. For this configuration, the **M** matrix for a multiferroic material with a cycloidal magnetic ordering becomes:

$$
\begin{pmatrix}
\varepsilon_\perp & 0 & 0 & 0 & \rho & 0 \\
0 & \varepsilon_\perp & 0 & 0 & 0 & 0 \\
0 & 0 & \varepsilon_\parallel & 0 & 0 & 0 \\
0 & 0 & 0 & \mu_\perp & 0 & 0 \\
\rho' & 0 & 0 & 0 & \mu_\perp & 0 \\
0 & 0 & 0 & 0 & 0 & \mu_\parallel
\end{pmatrix},
\tag{25}
$$

and its associated $\tilde{\Delta}$ matrix is calculated to be:

$$
\begin{pmatrix}
\rho' & \mu_\perp - \dfrac{N_0^2 \sin\left(\theta_0\right)^2}{\varepsilon_\parallel} & 0 & 0 \\
\varepsilon_\perp & \rho & 0 & 0 \\
0 & 0 & 0 & \mu_\perp \\
0 & 0 & \varepsilon_\perp - \dfrac{N_0^2 \sin\left(\theta_0\right)^2}{\mu_\parallel} & 0
\end{pmatrix}.
\tag{26}
$$

Inserting the $\tilde{\Delta}$ matrix from Eq. (26) into the Berreman equation [see Eq. (4)] returns the following four eigenvalue solutions:

$$q_1 = \frac{\omega}{2c}\left(\rho + \rho' + \sqrt{(\rho - \rho')^2 + 4\varepsilon_\perp \mu_\perp - \frac{4\varepsilon_\perp N_0^2 \sin(\theta_0)^2}{\varepsilon_\parallel}}\right)$$

$$q_2 = \frac{\omega}{c}\sqrt{\mu_\perp\left(\varepsilon_\perp - \frac{N_0^2 \sin(\theta_0)^2}{\mu_\parallel}\right)}$$

$$q_3 = \frac{\omega}{2c}\left(\rho + \rho' - \sqrt{(\rho - \rho')^2 + 4\varepsilon_\perp \mu_\perp - \frac{4\varepsilon_\perp N_0^2 \sin(\theta_0)^2}{\varepsilon_\parallel}}\right)$$

$$q_4 = -\frac{\omega}{c}\sqrt{\mu_\perp\left(\varepsilon_\perp - \frac{N_0^2 \sin(\theta_0)^2}{\mu_\parallel}\right)}$$

$$(27)$$

In Eq. (27), $q_1$ and $q_3$ are associated with $p$ polarized light and at normal incidence ($\theta_0 = 0$), these wave vectors reduce to exactly those derived in Ref. [10]. $q_1$ and $q_3$ represent forward and backward propagating waves, respectively. The wave vectors $q_2$ and $q_4$ are associated with $s$ polarized radiation and are similar in form to those derived for $s$ polarization in Case 1. Here $\varepsilon_{yy} = \varepsilon_{xx}$ as required to model uniaxial symmetry. With these derivations for the $z$ components, the complete description of the wave vectors for both polarization states of the forward propagating waves are:

$$\vec{k}_p = \left(\left(\frac{\omega}{c}\right)N_0\sin(\theta_0), 0, \frac{\omega}{2c}\left(\rho + \rho' + \sqrt{\varepsilon_\perp}\sqrt{\frac{(\rho - \rho')^2}{\varepsilon_\perp} + 4\mu_\perp - \frac{4N_0^2\sin(\theta_0)^2}{\varepsilon_\parallel}}\right)\right)$$

$$(28)$$

$$\vec{k}_s = \left(\left(\frac{\omega}{c}\right)N_0\sin(\theta_0), 0, \frac{\omega}{c}\sqrt{\mu_\perp}\sqrt{\left(\varepsilon_\perp - \frac{N_0^2\sin(\theta_0)^2}{\mu_\parallel}\right)}\right)$$

As evident from Eq. (28), this magneto-electric crystal will display birefringence as the two wave vectors will diverge in the direction of propagation [see Fig. 3(c)]. Of course, this result is expected for a uniaxial crystal. However, even if we had assumed isotropic $\hat{\varepsilon}$ and $\hat{\mu}$ tensors (which were not birefringent in Case 2), as it can be seen from Eq. (28), the birefringence would still have been in effect due to the presence of the magneto-electric tensors [see Fig. 3(d)]. With the definition $q_a = \sqrt{\left(\rho - \rho'\right)^2 + 4\varepsilon_\perp \mu_\perp - \dfrac{4\varepsilon_\perp N_0^2 \sin\left(\theta_0\right)^2}{\varepsilon_\parallel}}$, the associated eigenvectors (in columns) are:

$$
\begin{pmatrix}
1 & 0 & 1 & 0 \\
\dfrac{2\varepsilon_\perp}{\rho' - \rho + q_a} & 0 & \dfrac{2\varepsilon_\perp}{\rho' - \rho - q_a} & 0 \\
0 & 1 & 0 & 1 \\
0 & \dfrac{\sqrt{\varepsilon_\perp - \dfrac{N_0^2 \sin\left(\theta_0\right)^2}{\mu_\parallel}}}{\sqrt{\mu_\perp}} & 0 & -\dfrac{\sqrt{\varepsilon_\perp - \dfrac{N_0^2 \sin\left(\theta_0\right)^2}{\mu_\parallel}}}{\sqrt{\mu_\perp}}
\end{pmatrix}
\qquad (29)
$$

In Eq. (29), the first column is the eigenvector associated with $q_1$ while the second and fourth columns are associated with $q_2$. Solutions for waves, corresponding to these columns include factors $\exp\left(iq_1 z\right)$ and $\exp\left(\pm iq_2 z\right)$, respectively. The third column is associated with $q_3$ and the corresponding solution contains the factor $\exp\left(iq_3 z\right)$. As stated earlier, $q_{1,2}$ are the $z$ components of the wave vectors of the forward propagating waves while $q_{3,4}$ are the $z$ components of the wave vectors of the backward propagating waves. The eigenvectors in columns one and three are influenced by the magneto-electric effect. The forward propagating eigenvectors when combined with the tangential boundary conditions for $\vec{E}$ and $\vec{H}$ return the complex reflection coefficients which make up the Jones matrix:

$$\begin{pmatrix} \dfrac{2\varepsilon_{\perp}\cos(\theta_0) - N_0(\rho' - \rho + q_a)}{2\varepsilon_{\perp}\cos(\theta_0) + N_0(\rho' - \rho + q_a)} & 0 \\ & \\ 0 & \dfrac{N_0\sqrt{\mu_{\perp}}\cos(\theta_0) - \sqrt{\varepsilon_{\perp} - \dfrac{N_0^2\sin(\theta_0)^2}{\mu_{\parallel}}}}{N_0\sqrt{\mu_{\perp}}\cos(\theta_0) + \sqrt{\varepsilon_{\perp} - \dfrac{N_0^2\sin(\theta_0)^2}{\mu_{\parallel}}}} \end{pmatrix} \quad (30)$$

The absence of the cross polarization terms on the off-diagonal is due to the coincidence of the crystal principal axes with the laboratory system and also due to the location of $\rho_{xy}$ in the optical matrix. This term couples with $H_y$ to provide an additional influence on $E_x$. However, both $E_x$ and $H_y$ are the complementary vectors for $p$ polarization and therefore, no cross polarization terms arise. In the next section, we will show how the position of the $\hat{\rho}$ in the optical matrix can give rise to the off-diagonal terms of the Jones matrix. As will be discussed later, with proper dispersion relations for the $\hat{\varepsilon}, \hat{\mu}$ and $\hat{\rho}$ tensors, the reflectivity spectra for this crystal can be simulated using Eq. (30).

The presence of three distinct eigenvalues for $\tilde{\Delta}$ matrix gives rise to the interesting phenomenon of the irreversibility of the $p$ polarized wave propagation. The wave propagation associated with wave vectors $q_1$ and $q_3$ is irreversible because they have different phase velocities.[10] For oblique AOI, the same path will not be followed for each the forward and backward propagating waves. On the other hand, $q_2$ and $q_4$ are clearly reversible because the electro-magnetic effect is not picked up for $s$ polarized incident radiation. It also shows that optical reflectivity spectra measured for $s$ polarized radiation are not sensitive to magneto-electric excitations. As previously explained, the $s$ polarized eigenvalues are entirely consistent with those of Case 1 after adjusting for uniaxial symmetry.

As in the previous two cases, through $4 \times 4$ matrix formalism, comparisons can be made between the direction of propagation and the direction of energy flow described by Poynting vector. For this crystal symmetry, the formulas for the recapture of the $z$ components of $\vec{E}$ and $\vec{H}$ vectors are identical to those in Eq. (12) derived in Case 1. For wave corresponding to the eigenvalue $q_1$ , $\vec{E}$ and $\vec{H}$ are:

$$\vec{E} = E_x \left( 1, 0, \frac{-2\varepsilon_\perp N_0 \sin(\theta_0)}{\varepsilon_\parallel (\rho' - \rho + q_a)} \right) e^{iq_1 z}$$

$$\vec{H} = E_x \left( 0, \frac{2\varepsilon_\perp}{\rho' - \rho + q_a}, 0 \right) e^{iq_1 z}$$

(31)

and the associated Poynting vector is calculated to be :

$$\vec{S}_{q1} = \left( \left| \frac{\varepsilon_\perp}{(\rho - \rho' + q_a)} \right|^2 \frac{2 N_0 \sin(\theta_0)}{\varepsilon_\parallel}, 0, \left( \frac{\varepsilon_\perp}{\rho' - \rho + q_a} \right)^* \right)$$

(32)

From Eq. (32), the tangent of the S vector angle in the medium is:

$$\tan(\theta_S) = \frac{2\varepsilon_\perp N_0 \sin(\theta_0)}{\varepsilon_\parallel (\rho' - \rho + q_a)}$$

(33)

Similarly, from Eq. (28), the tangent of the k vector angle in the medium is:

$$\tan\left(\theta_k\right) = \frac{2N_0 \sin\left(\theta_0\right)}{\left(\rho + \rho' + q_a\right)} \tag{34}$$

Again, while Eq. (33) and Eq. (34) are similar in form, they are not identical. As a result, there will be a divergence between the direction of wave propagation $\vec{k}$ and the direction of energy flow $\vec{S}$. As was shown in Case 1, for the $s$ polarization state, the $\vec{k}$ and $\vec{S}$ vectors are also not coincident. This configuration is simulated in Fig. 3(c) for a material with diagonal tensors: $\varepsilon = \left(4,4,5\right)$, $\mu = \left(2,2,3\right)$, and $\rho_{xy} = 3$. Again, all four vectors are distinct.

Finally, we will briefly address the wave whose $z$ component is given by $q_3$. Following the same procedure for $q_1$, the wave vector and Poynting vector are calculated to be:

$$\vec{k}_{q3} = \left(\left(\frac{\omega}{c}\right)N_0 \sin\left(\theta_0\right), 0, \frac{\omega}{2c}\left(\rho + \rho' - \sqrt{\varepsilon_\perp}\sqrt{\frac{\left(\rho - \rho'\right)^2}{\varepsilon_\perp} + 4\mu_\perp - \frac{4N_0^2 \sin\left(\theta_0\right)^2}{\varepsilon_\parallel}}\right)\right)$$

$$\vec{S}_{q3} = \left(\left|\frac{\varepsilon_\perp}{\left(\rho - \rho' - q_a\right)}\right|^2 \frac{2N_0 \sin\left(\theta_0\right)}{\varepsilon_\parallel}, 0, \left(\frac{\varepsilon_\perp}{\rho' - \rho - q_a}\right)^*\right)$$

(35)

From Eq.(35), it can be shown that the $z$ component of $\vec{k}_{q3}$ becomes zero at normal incidence when $\rho\rho' = \varepsilon_\perp\mu_\perp$. This suggests that, under this condition, this wave vector will not propagate in the medium. We note the similarity between this condition and Dzyaloshinsky's inequality for the magneto-electric tensor in a static case, $\alpha \cdot \alpha' \le \varepsilon \cdot \mu$, which is based on the Gibbs Free Energy consideration.[11, 12, 51]

### D. Case 4-Isotropic $\hat{\varepsilon}$ and $\hat{\mu}$ Tensors; Isotropic Magneto-Electric Tensors

In Case 4, the constitutive relationships are described by simultaneously diagonalized isotropic tensors. While this configuration is not strictly allowed given symmetry constraints, for certain multiferroics in a polycrystalline form, the macroscopic anisotropy is small and the crystal can be effectively modeled using the isotropic assumption. The corresponding optical matrix **M** is:

$$
\begin{pmatrix}
\varepsilon & 0 & 0 & \rho & 0 & 0 \\
0 & \varepsilon & 0 & 0 & \rho & 0 \\
0 & 0 & \varepsilon & 0 & 0 & \rho \\
\rho' & 0 & 0 & \mu & 0 & 0 \\
0 & \rho' & 0 & 0 & \mu & 0 \\
0 & 0 & \rho' & 0 & 0 & \mu
\end{pmatrix}
\tag{36}
$$

The associated $\tilde{\Delta}$ matrix is:

$$
\tilde{\Delta} =
\begin{pmatrix}
0 & \mu - \dfrac{\mu N_0^2 \sin(\theta_0)^2}{\varepsilon\mu - \rho\rho'} & \rho' - \dfrac{\rho N_0^2 \sin(\theta_0)^2}{\varepsilon\mu - \rho\rho'} & 0 \\
\varepsilon & 0 & 0 & -\rho \\
\rho' & 0 & 0 & \mu \\
0 & \rho - \dfrac{\rho' N_0^2 \sin(\theta_0)^2}{\varepsilon\mu - \rho\rho'} & \varepsilon - \dfrac{\varepsilon N_0^2 \sin(\theta_0)^2}{\varepsilon\mu - \rho\rho'} & 0
\end{pmatrix}
\tag{37}
$$

which has the following four eigenvalue solutions:

$$q_{z1,z3} = \pm \frac{\omega}{c} \sqrt{\frac{2\varepsilon\mu - (\rho^2 + \rho'^2) + (\rho - \rho')K - 2N_0^2 \sin^2(\theta_0)}{2}}$$

$$q_{z2,z4} = \pm \frac{\omega}{c} \sqrt{\frac{2\varepsilon\mu - (\rho^2 + \rho'^2) - (\rho - \rho')K - 2N_0^2 \sin^2(\theta_0)}{2}}$$

$$(38)$$

In Eq. (38), $K = \sqrt{(\rho + \rho')^2 - 4\varepsilon\mu}$. The two $\vec{k}$ vectors in the medium are $\left( \frac{\omega}{c} N_0 \sin(\theta_0), 0, q_{z1} \right)$

and $\left( \frac{\omega}{c} N_0 \sin(\theta_0), 0, q_{z2} \right)$. For $\rho = \rho'$ the wave vectors are identical and there will be no birefringence. However, for $\rho \neq \rho'$, which is possible in the dynamic case and/or in the medium with chirality, $q_{z1} \neq q_{z2}$ and there will be two refracted waves with the direction of each wave being influenced by the combination of the $\varepsilon$, $\mu$, $\rho$ and $\rho'$ parameters. Such a material is bi-anisotropic and behaves similar to a birefringent medium. For $\rho \neq \rho'$, the magneto-electric tensors introduce birefringence even in the presence of isotropic $\hat{\varepsilon}$ and $\hat{\mu}$. In Eq. (38), the positive signs indicate forward propagating waves while the negative signs indicate backward propagating waves. Note that the phase is identical for both forward and backward propagating waves. Reversibility for this configuration is expected, of course, given that all tensors are isotropic. For $\rho \neq \rho'$, the eigenvector solutions $\left( [E_x, H_y, E_y, -H_x]^T \right)$, in columns, for the $\tilde{\Delta}$

matrix are:

$$\begin{pmatrix} 1 & 1 & 1 & 1 \\ \dfrac{-c\Xi L q_{z1}}{\omega\mu\left[ \Xi M + Q N_0^2 \sin^2(\theta_0) \right]} & \dfrac{c\Xi L q_{z1}}{\omega\mu\left[ \Xi M + Q N_0^2 \sin^2(\theta_0) \right]} & \dfrac{c\Xi U q_{z2}}{\omega\mu\left[ \Xi Q + M N_0^2 \sin^2(\theta_0) \right]} & \dfrac{-c\Xi U q_{z2}}{\omega\mu\left[ \Xi Q + M N_0^2 \sin^2(\theta_0) \right]} \\ \dfrac{2c\Xi q_{z1}}{\omega\left[ \Xi M + Q N_0^2 \sin^2(\theta_0) \right]} & \dfrac{-2c\Xi q_{z1}}{\omega\left[ \Xi M + Q N_0^2 \sin^2(\theta_0) \right]} & \dfrac{-2c\Xi q_{z2}}{\omega\left[ \Xi Q + M N_0^2 \sin^2(\theta_0) \right]} & \dfrac{2c\Xi q_{z2}}{\omega\left[ \Xi Q + M N_0^2 \sin^2(\theta_0) \right]} \\ \dfrac{L}{2\mu} & \dfrac{L}{2\mu} & \dfrac{U}{2\mu} & \dfrac{U}{2\mu} \end{pmatrix}$$

$$(39)$$

where

$$\Xi = \rho\rho' - \varepsilon\mu$$

$$K = \sqrt{\left(\rho + \rho'\right)^2 - 4\varepsilon\mu}$$

$$L = \rho + \rho' - K$$

$$M = -\rho + \rho' + K \qquad (40)$$

$$Q = \rho - \rho' + K$$

$$U = \rho + \rho' + K$$

As can be seen in Eq. (39), there are four distinct eigenvector solutions. The first two columns of Eq. (39) are associated with the $q_{z1}$ eigenvalue and their complete description requires multiplication by $\exp\left(\pm i q_{z1} z\right)$. The last two columns are associated with the $q_{z2}$ eigenvalue and their complete description requires multiplication by $\exp\left(\pm i q_{z2} z\right)$. Completing the reflection calculation using $4 \times 4$ matrix formalism returns four complex reflection coefficients:

$$r_{pp} = -\frac{2q_{z1}q_{z2}K + 2k_0^2 K(\rho\rho' - \varepsilon\mu) - k_0(\varepsilon - \mu)\left[q_{z1}(\rho - \rho' + K) + q_{z2}(-\rho + \rho' + K)\right]}{2q_{z1}q_{z2}K + 2k_0^2 K(-\rho\rho' + \varepsilon\mu) + k_0(\varepsilon + \mu)\left[q_{z1}(\rho - \rho' + K) + q_{z2}(-\rho + \rho' + K)\right]}$$

$$r_{ps} = \frac{2k_0\left\{-q_{z1}\rho K - q_{z2}\rho K - (q_{z1} - q_{z2})\left[\rho(\rho + \rho') - 2\varepsilon\mu\right]\right\}}{2q_{z1}q_{z2}K + 2k_0^2 K(-\rho\rho' + \varepsilon\mu) + k_0(\varepsilon + \mu)\left[q_{z1}(\rho - \rho' + K) + q_{z2}(-\rho + \rho' + K)\right]}$$

$$r_{sp} = \frac{2k_0\left\{-q_{z1}\rho' K - q_{z2}\rho' K + (q_{z1} - q_{z2})\left[\rho'(\rho + \rho') - 2\varepsilon\mu\right]\right\}}{2q_{z1}q_{z2}K + 2k_0^2 K(-\rho\rho' + \varepsilon\mu) + k_0(\varepsilon + \mu)\left[q_{z1}(\rho - \rho' + K) + q_{z2}(-\rho + \rho' + K)\right]}$$

$$r_{ss} = -\frac{2q_{z1}q_{z2}K + 2k_0^2 K(\rho\rho' - \varepsilon\mu) + k_0(\varepsilon - \mu)\left[q_{z1}(\rho - \rho' + K) + q_{z2}(-\rho + \rho' + K)\right]}{2q_{z1}q_{z2}K + 2k_0^2 K(-\rho\rho' + \varepsilon\mu) + k_0(\varepsilon + \mu)(q_{z1}(\rho - \rho' + K) + q_{z2}(-\rho + \rho' + K))}$$

$$(41)$$

In contrast to the previous cases, we see that the off-diagonal Jones matrix elements are not necessarily zero. As expected, if $\rho$ and $\rho'$ are identically zero, the off-diagonal elements vanish and $r_{pp}$ and $r_{ss}$ reduce to previously calculated results for the $p$ and $s$ polarized reflection coefficients for a non magneto-electric semi-infinite medium [48]. It is important to note that Eq. (41) is written in terms of $\hat{\rho}$ and $\hat{\rho}'$, the general magneto-electric tensors which, by definition, can incorporate the magneto-electric effect, chirality or both. It has been derived using $\vec{E}$ and $\vec{H}$ as the base field vectors.[8, 13, 14] Other conventions are possible, for example, using $\vec{E}$ and $\vec{B}$ as the base field vectors.[39, 40] When analyzing chirality only ($\hat{\rho} = \hat{\xi}$), Eq. (41) produces numerical results consistent with Ref. [39] for $\xi^2 \ll \varepsilon\mu$ (when the $\left(\vec{E}, \vec{H}\right)$ basis is used). When analyzing both the magneto-electric effect and chirality at normal incidence ($\hat{\rho} = \hat{\alpha} + i\hat{\xi}$), Eq. (41) produces numerical results consistent with the Berreman formalism and with Ref. [8]. In Eq. (41), we use sign conventions derived from a chiral only base case: the ellipsometric parameter $\Delta$ is equal to $180^0$ for the diagonal terms; for the off-diagonal terms $r_{sp} = -r_{ps}$. We note that this convention is not universally applied in the literature.

For this symmetry, the formulas for recapture of the $z$ components of $\vec{E}$ and $\vec{H}$ vectors are more complicated than for the previous cases and are given by:

$$E_z = N_0 \sin\left(\theta_0\right)\left(\frac{E_y\rho + H_y\mu}{\rho\rho' - \varepsilon\mu}\right)$$

$$H_z = -N_0 \sin\left(\theta_0\right)\left(\frac{H_y\rho' + E_y\varepsilon}{\rho\rho' - \varepsilon\mu}\right)$$

(42)

We again note that the solutions in Eq. (42) can be applied to both eigenvectors. For purposes of illustration, we will analyze propagation associated with the $q_{z1}$ eigenvalue only. For a wave associated with $q_{z1}$, the electromagnetic fields in the medium are:

$$\vec{E} = \left( 1, \frac{2c(\rho\rho' - \varepsilon\mu)q_{z1}}{\omega\left[(\rho\rho' - \varepsilon\mu)M + QN_0^2\sin^2(\theta_0)\right]}, N_0\sin(\theta_0)\frac{c}{\omega}\frac{q_{z1}(2\rho - L)}{\left[(\rho\rho' - \varepsilon\mu)M + QN_0^2\sin^2(\theta_0)\right]} \right)e^{iq_{z1}z}$$

$$\vec{H} = \left( -\frac{L}{2\mu}, \frac{-c(\rho\rho' - \varepsilon\mu)Lq_{z1}}{\omega\mu\left[(\rho\rho' - \varepsilon\mu)M + QN_0^2\sin^2(\theta_0)\right]}, N_0\sin(\theta_0)\frac{c}{\omega}\frac{+q_{z1}(L\rho' - 2\varepsilon\mu)}{\mu\left[(\rho\rho' - \varepsilon\mu)M + QN_0^2\sin^2(\theta_0)\right]} \right)e^{iq_{z1}z}$$

$$(43)$$

The $y$ and $z$ component terms in Eq. (43) are dependent upon the angle of incidence $\theta_0$. Therefore, even at normal incidence, the eigenvector solutions will have both $x$ and $y$ components for each of $\vec{E}$ and $\vec{H}$. Accordingly, while still vibrating in the x-y plane, the magneto-electric effect causes the eigenvector to be rotated off the principal axes as it propagates into the material. This is different from orthorhombic $\hat{\varepsilon}$ and $\hat{\mu}$ (Case 1), for example, where the eigenvectors remain on the principal axes only [see Eq. (9)]. For general AOI, the calculation of the Poynting vector, $\vec{S} = \frac{1}{2}\vec{E} \times \vec{H}^*$, is a complicated algebraic expression and proper modeling requires a numerical approach. Analytically, it can be shown that at normal incidence ($\theta_0 = 0$), both the $\vec{k}$ and $\vec{S}$ vectors are parallel with propagation along the $z$ axis only, as expected for this trivial case. Numerical simulations show that for variable AOI, the two Poynting vectors are coincident with the two $\vec{k}$ vectors. We ascribe this to the isotropic symmetry assumption for the $\rho$ tensor. We simulate this material having diagonal tensors: $\varepsilon = (4,4,4)$ and $\mu = (2,2,2)$ and real values of $\rho = (2.5, 2.5, 2.5)$. Fig. 4(a) illustrates that all four vectors are coincident. However, if the medium is chiral ($\rho$ is complex), it can be shown that birefringence will emerge.

Finally, we address the interesting condition when $\rho\rho' = \varepsilon\mu$. This configuration is modeled for an oblique AOI in Fig. 4(b) for a material with diagonal tensors: $\varepsilon = (4, 4, 4)$ and $\mu = (2, 2, 2)$ and $\rho = (\sqrt{8}, \sqrt{8}, \sqrt{8})$. At oblique angles, the $\vec{k}$ vector retains a non-zero $x$ component. However, all four eigenvalues are purely imaginary which is suggestive of a decaying wave in the $z$ direction. In fact, at normal incidence, $q_{z1}$ and $q_{z3}$ are identically zero suggesting that the wave vector will not penetrate the medium. Interestingly, this condition, derived using $4 \times 4$ matrix formalism, coincides with the thermodynamically derived limitation that the square of the magneto-electric susceptibility must be less than the geometric mean of the diagonalized $\hat{\varepsilon}$ and $\hat{\mu}$ tensors.[52, 53] We note further that the condition $\rho\rho' = \varepsilon\mu$ is identical to the condition in Case 3 which was derived using a completely different symmetry for the $\hat{\rho}$ tensor.

### E. Case 5-Anisotropic $\hat{\varepsilon}$ and $\hat{\mu}$ Tensors; Anisotropic Magneto-Electric Tensors

In Case 5, the configuration of anisotropic $\hat{\varepsilon}$, $\hat{\mu}$ and magneto-electric tensors is examined. Orthorhombic symmetry, which is appropriate for crystals belonging to the 222 point group, is chosen for each tensor.[9, 11] The **M** matrix for this configuration is:

$$\begin{pmatrix} \varepsilon_{xx} & 0 & 0 & \rho_{xx} & 0 & 0 \\ 0 & \varepsilon_{yy} & 0 & 0 & \rho_{yy} & 0 \\ 0 & 0 & \varepsilon_{zz} & 0 & 0 & \rho_{zz} \\ \rho'_{xx} & 0 & 0 & \mu_{xx} & 0 & 0 \\ 0 & \rho'_{yy} & 0 & 0 & \mu_{yy} & 0 \\ 0 & 0 & \rho'_{zz} & 0 & 0 & \mu_{zz} \end{pmatrix}, \tag{44}$$

and its associated $\tilde{\Delta}$ matrix is calculated as:

$$\tilde{\Delta} = \begin{pmatrix} 0 & \mu_{yy} - \dfrac{\mu_{zz} N_0^{\ 2} \sin(\theta_0)^2}{\varepsilon_{zz}\mu_{zz} - \rho_{zz}\rho'_{zz}} & \rho'_{yy} - \dfrac{\rho_{zz} N_0^{\ 2} \sin(\theta_0)^2}{\varepsilon_{zz}\mu_{zz} - \rho_{zz}\rho'_{zz}} & 0 \\ \varepsilon_{xx} & 0 & 0 & -\rho_{xx} \\ \rho'_{xx} & 0 & 0 & \mu_{xx} \\ 0 & \rho_{yy} - \dfrac{\rho'_{zz} N_0^{\ 2} \sin(\theta_0)^2}{\varepsilon_{zz}\mu_{zz} - \rho_{zz}\rho'_{zz}} & \varepsilon_{yy} - \dfrac{\varepsilon_{zz} N_0^{\ 2} \sin(\theta_0)^2}{\varepsilon_{zz}\mu_{zz} - \rho_{zz}\rho'_{zz}} & 0 \end{pmatrix} \tag{45}$$

In Eq. (45), it can be seen that all direction components of each tensor enter into the $\tilde{\Delta}$ matrix. With 12 different variables entering into the calculation, the analytic solution for the wave vectors is quite complicated. Accordingly, this is an example of a configuration which requires numerical analysis for proper modeling. This configuration is simulated in Fig. 4(c) for a material with diagonal tensors: $\varepsilon = (4,6,8)$ and $\mu = (3,4,5)$ and $\rho = (1,2,3)$. As expected, all four vectors are distinct. By examining the denominators in the first and fourth rows of the $\tilde{\Delta}$ matrix in Eq. (45) we see that the limiting condition for an anisotropic magneto-electric material is given as $\varepsilon_{zz}\mu_{zz} = \rho_{zz}\rho'_{zz}$.

## III. THIN FILM CONFIGURATION

The analytical procedures for thin films using $4 \times 4$ matrix formalism are identical to those of the semi-infinite configuration up to the calculation of the complex reflection and transmission coefficients. In the following analysis, we restrict our work to a single layer thin film structure with the thickness $d$. For this configuration, both forward and backward propagating waves (*i.e.*, all four eigenvectors) are needed to satisfy the electromagnetic boundary conditions at both top and bottom interfaces. The tangential components of the electric and magnetic field vectors are matched at $z = 0$ and $z = d$ to produce two generalized field vectors $\psi(0)$ and $\psi(d)$, respectively. A thin film layer matrix $L$ is utilized to relate the fields inside the film of thickness $d$ at its two boundaries:[14, 15]

$$\psi(d) = L\psi(0) \tag{46}$$

$L$ is a $4 \times 4$ matrix calculated from the eigenvalues and eigenvectors of the $\tilde{\Delta}$ matrix according to:

$$L(d) = \tilde{\Psi} * K(d) * \tilde{\Psi}^{-1} \tag{47}$$

In Eq. (47), $\tilde{\Psi}$ is composed of the four $\tilde{\Delta}$ eigenvectors as columns while $K$ is a diagonal matrix given by $K_{ll} = \exp(iq_l d)$ with $q_l$ representing the four eigenvalues of $\tilde{\Delta}$. After some algebra relating the incident and reflected waves, the complex reflection and transmission coefficients for a thin film can be calculated.

### A. Case 1- Anisotropic $\hat{\varepsilon}$ and $\hat{\mu}$ Tensors; No Magneto-Electric Activity

Analytic expressions for the Jones matrix components for the case of orthorhombic $\hat{\varepsilon}$ and $\hat{\mu}$ for both $p$ and $s$ polarizations have been given in Ref. [48]. For purposes of comparison to other material symmetries, the equations are reproduced here. The complex reflection and transmission coefficients for *p* polarized radiation are:

$$r_{pp} = \frac{q_{zp}\cos(q_{zp}d)\left(\dfrac{N_2}{N_0}k_{z0} - \dfrac{N_0}{N_2}k_{z2}\right) + i\left(\dfrac{N_0 N_2 q_{zp}^2}{\varepsilon_{xx}} - \dfrac{\varepsilon_{xx}k_{z0}k_{z2}}{N_0 N_2}\right)\sin(q_{zp}d)}{q_{zp}\cos(q_{zp}d)\left(\dfrac{N_2}{N_0}k_{z0} + \dfrac{N_0}{N_2}k_{z2}\right) - i\left(\dfrac{N_0 N_2 q_{zp}^2}{\varepsilon_{xx}} + \dfrac{\varepsilon_{xx}k_{z0}k_{z2}}{N_0 N_2}\right)\sin(q_{zp}d)}$$

$$t_{pp} = \frac{2k_{z0}q_{zp}}{q_{zp}\cos(q_{zp}d)\left(\dfrac{N_2}{N_0}k_{z0} + \dfrac{N_0}{N_2}k_{z2}\right) - i\left(\dfrac{N_0 N_2 q_{zp}^2}{\varepsilon_{xx}} + \dfrac{\varepsilon_{xx}k_{z0}k_{z2}}{N_0 N_2}\right)\sin(q_{zp}d)}$$

(48)

The complex reflection and transmission coefficients for *s* polarized radiation are:

$$r_{ss} = \frac{q_{zs}\cos(q_{zs}d)\left(k_{z0} - k_{z2}\right) + i\left(\dfrac{q_{zs}^2}{\mu_{xx}} - k_{z0}k_{z2}\mu_{xx}\right)\sin(q_{zs}d)}{q_{zs}\cos(q_{zs}d)\left(k_{z0} + k_{z2}\right) - i\left(\dfrac{q_{zs}^2}{\mu_{xx}} + k_{z0}k_{z2}\mu_{xx}\right)\sin(q_{zs}d)}$$

$$t_{ss} = \frac{2k_{z0}q_{zs}}{q_{zs}\cos(q_{zs}d)\left(k_{z0} + k_{z2}\right) - i\left(\dfrac{q_{zs}^2}{\mu_{xx}} + k_{z0}k_{z2}\mu_{xx}\right)\sin(q_{zs}d)}$$

(49)

In Eq. (48) and Eq. (49), $q_{zp}$ and $q_{zs}$ have the same definitions as derived in Case 1 for the semi-infinite configuration [see Eq. (8)]. $k_{z0} = \dfrac{\omega}{c}N_0\cos(\theta_0)$ and $k_{z2} = \dfrac{\omega}{c}N_2\cos(\theta_2)$ are the $z$ components of the incident and substrate wave vectors, respectively.

## B. Case 3-Anisotropic $\varepsilon$ and $\mu$ Tensors; Off-Diagonal Magneto-Electric Tensors

In this section, we will analyze the thin film complex reflection and transmission coefficients of the crystal discussed in Case 3 of Section 2. This work permits the interesting analysis of the impact of the magneto-electric tensor on reflectivity and transmission. We will show that the magneto-electric tensor affects only the $p$ polarization terms since it is only this polarization that has been influenced by the magneto-electric effect for this particular symmetry [see Eq. (27)]. Using the above procedures for $4 \times 4$ matrix formalism, the complex reflection coefficients are calculated to be:

$$r_{pp} = \frac{\frac{\omega}{c} q_a \cos\left(\frac{\omega}{2c} q_a d\right)\left(\frac{N_2}{N_0} k_{z0} - \frac{N_0}{N_2} k_{z2}\right) - i \frac{\omega}{c}(\rho - \rho')\left(\frac{N_2}{N_0} k_{z0} + \frac{N_0}{N_2} k_{z2}\right)\sin\left(\frac{\omega}{2c} q_a d\right) + 2i\left(\frac{N_0 N_2 q_\infty^2}{\varepsilon_{xx}} - \frac{\varepsilon_{xx} k_{z0} k_{z2}}{N_0 N_2}\right)\sin\left(\frac{\omega}{2c} q_a d\right)}{\frac{\omega}{c} q_a \cos\left(\frac{\omega}{2c} q_a d\right)\left(\frac{N_2}{N_0} k_{z0} + \frac{N_0}{N_2} k_{z2}\right) + i \frac{\omega}{c}(\rho - \rho')\left(\frac{N_0}{N_2} k_{z2} - \frac{N_2}{N_0} k_{z0}\right)\sin\left(\frac{\omega}{2c} q_a d\right) - 2i\left(\frac{N_0 N_2 q_\infty^2}{\varepsilon_{xx}} + \frac{\varepsilon_{xx} k_{z0} k_{z2}}{N_0 N_2}\right)\sin\left(\frac{\omega}{2c} q_a d\right)}$$

$$r_{ss} = \frac{q_{zs}\cos(q_{zs}d)\left(k_{z0} - k_{z2}\right) + i\left(\frac{q_{zs}^2}{\mu_{xx}} - k_{z0} k_{z2}\mu_{xx}\right)\sin(q_{zs}d)}{q_{zs}\cos(q_{zs}d)\left(k_{z0} + k_{z2}\right) - i\left(\frac{q_{zs}^2}{\mu_{xx}} + k_{z0} k_{z2}\mu_{xx}\right)\sin(q_{zs}d)}$$

$$(50)$$

The formulas for the complex transmission coefficients are:

$$t_{pp} = \frac{2e^{i\frac{\omega}{c}\left(\frac{\rho+\rho'}{2}\right)}\frac{\omega}{c}q_a k_{z0}}{\frac{\omega}{c}q_a \cos\left(\frac{\omega}{2c}q_a d\right)\left(\frac{N_2}{N_0}k_{z0}+\frac{N_0}{N_2}k_{z2}\right)+i\frac{\omega}{c}(\rho-\rho')\left(\frac{N_0}{N_2}k_{z2}-\frac{N_2}{N_0}k_{z0}\right)\sin\left(\frac{\omega}{2c}q_a d\right)-2i\left(\frac{N_0 N_2 q_{zx}^2}{\varepsilon_{xx}}+\frac{\varepsilon_{xx}k_{z0}k_{z2}}{N_0 N_2}\right)\sin\left(\frac{\omega}{2c}q_a d\right)}$$

$$t_{ss} = \frac{2k_{z0}q_{zs}}{q_{zs}\cos(q_{zs}d)(k_{z0}+k_{z2})-i\left(\frac{q_{zs}^2}{\mu_{xx}}+k_{z0}k_{z2}\mu_{xx}\right)\sin(q_{zs}d)}$$

$$(51)$$

As for the semi-infinite configuration, the off diagonal Jones matrix elements for the thin film configuration vanish due to the coincidence of the material's principal axes with the laboratory system in the above derivation and also due to the location of the magneto-electric tensor element in the optical matrix.

As expected, the magneto-electric tensors affect only the $p$ polarization terms. The equations for $r_{ss}$ and $t_{ss}$ are the same as for Case 1 since the $s$ polarization is not affected. In Eq. (50) and Eq. (51), $q_a = \sqrt{(\rho-\rho')^2 + 4\varepsilon_\perp \mu_\perp - \frac{4\varepsilon_\perp N_0^2 \sin(\theta_0)^2}{\varepsilon_\parallel}}$. It is interesting to note that it is only $q_a$ and not the entire eigenvalue expression for $q_1$ [see Eq. (27)] that enters into the argument for the trigonometric functions in both the thin film reflection and transmission coefficients. In $r_{pp}$, the magneto-electric terms enter as the middle terms of each of the numerator and denominator in Eq. (50) and for transmission they enter in the middle term of the denominator in $t_{pp}$. If $\rho = \rho'$, it can be seen that $\frac{\omega}{c}q_a = 2q_{zp}$, where $q_{zp}$ is as defined in Case 1 for the semi-infinite configuration. Under this condition, the magneto-electric terms vanish and $r_{pp}$ and $t_{pp}$ reduce to the identical expressions derived in Case 1 for thin films. We note further that this scenario is also consistent with the fact that the magneto-electric effect for hexagonal manganites is forbidden for symmetry reasons for the static case.[54] Eq. (50) and Eq. (51) , which accommodate variable AOI, should be of significant use to experimentalists in the analysis of the reflectivity spectra of magneto-electric thin film materials and should also assist in the proper characterization of the magneto-electric tensor and in the study of electromagnons. We note that if the experiment is designed as a vacuum–thin film–vacuum configuration, then the first term in the numerators for both reflection polarizations vanish and the formulas are further simplified:

$$r_{pp} = \frac{i\left(\dfrac{q_{zp}^2}{k_{z0}\varepsilon_{xx}} - \varepsilon_{xx}k_{z0}\right)\sin\left(\dfrac{\omega}{2c}q_a d\right) - i\dfrac{\omega}{c}(\rho - \rho')\sin\left(\dfrac{\omega}{2c}q_a d\right)}{\dfrac{\omega}{c}q_a \cos\left(\dfrac{\omega}{2c}q_a d\right) - i\left(\dfrac{q_{zp}^2}{k_{z0}\varepsilon_{xx}} + \varepsilon_{xx}k_{z0}\right)\sin\left(\dfrac{\omega}{2c}q_a d\right)}$$

(52)

$$r_{ss} = \frac{\dfrac{i}{2}\left(\dfrac{q_{zs}}{k_{z0}\mu_{xx}} - \dfrac{k_{z0}\mu_{xx}}{q_{zs}}\right)\sin(q_{zs}d)}{\cos(q_{zs}d) - \dfrac{i}{2}\left(\dfrac{q_{zs}}{k_{z0}\mu_{xx}} + \dfrac{k_{z0}\mu_{xx}}{q_{zs}}\right)\sin(q_{zs}d)}$$

## 4. DISPERSION MODELS FOR $\varepsilon$ AND $\mu$

In order to simulate the response functions in the optical spectra, such as, for example, normal incidence reflectivity $R_s(\omega)$ and transmittivity $T_s(\omega)$, certain assumptions must be made about the dispersion models. A common approach is to model the elementary excitations using a combination of Lorentzian oscillators. We first consider models for $\hat{\varepsilon}$ and $\hat{\mu}$ :

$$\varepsilon(\omega) = \varepsilon_\infty + \sum_{j=1}^{N} \frac{S_{j,e}\omega_{j,e0}{}^2}{(\omega_{j,e0}{}^2 - \omega^2 - i\gamma_{j,e}\omega)}$$

$$\mu(\omega) = \mu_\infty + \sum_{j=1}^{M} \frac{S_{j,m}\omega_{j,m0}{}^2}{(\omega_{j,m0}{}^2 - \omega^2 - i\gamma_{j,m}\omega)}, \tag{53}$$

where $\varepsilon_\infty$ and $\mu_\infty$ are the infinite-frequency values of the dielectric function ($\mu_\infty \cong 1$), $S_{e,m}$ is the oscillator strength, $\gamma_{e,m}$ is the damping constant, and $\omega_{e0,m0}$ is the resonance frequency of the excitation. Poles in the Lorentzian formulas are also known as modes for the response functions. We note in Eq. (53) that for metamaterials, the model for magnetic permeability is adjusted from the Lorentzian model via the replacement of $\omega_{m0}^2$ with $\omega^2$ in the numerator. This is known as the Pendry model and it ensures that the static value for the magnetic permeability is equal to 1. For multiferroics, this condition is not applicable and we use the SHO model. Other dispersion models, including the Coupled Harmonic Oscillator (CHO) model[17], can be also used to describe the response functions, but they are not considered in this paper.

With the above dispersion formulas, expressions for reflected and transmitted intensities can be obtained by multiplying the complex formulas by their complex conjugate. For example, for thin film $s$ polarization:

$$R_s(\omega) = r_{ss} \times r_{ss}{}^*$$

$$T_s(\omega) = t_{ss} \times t_{ss}{}^*} \tag{54}$$

It is clear from the foregoing analysis that the intensities in Eq. (54) are functions of $\hat{\varepsilon}$, $\hat{\mu}$, $\hat{\rho}$ and $\hat{\rho}'$.

From the dispersion formulas alone, it is possible to make conclusions about two interesting optical effects possible for complex media: (i) the inverted Lorentzian shape in reflectivity spectra for a pure magnetic dipole excitation; and (ii) the phenomenon of negative index of refraction (NIR).

As explained in Ref. [47], the shape of the response function of a pure magnetic dipole is best understood by using the Veselago approach presented in Section 2 B, where $n \to \sqrt{\varepsilon / \mu}$. It is assumed that the natural frequency of the magnetic dipole is far from dielectric resonance so that the dielectric function can be treated as a constant, $\varepsilon_\infty$. That is, we assume $S_e = 0$. The expression for reflection using the Veselago approach becomes:[47]

$$R_{ss}(\omega) \approx f\left( \sqrt{\varepsilon_\infty - \frac{\varepsilon_\infty S_m \omega_m^2}{(\omega_m^2 - \omega^2 - i\gamma\omega)}} \right), \qquad (55)$$

where $f(x) = \left| (1-x)/(1+x) \right|^2$. The negative sign in Eq. (55) corresponds to the inverted Lorentzian shape of a pure magnetic dipole with an adjusted oscillator strength (AOS) $S_R = S_m \cdot \varepsilon_\infty$ and the resonance frequency $\omega_m = \omega_{m,0} \cdot \sqrt{1 + S_m}$. As will be discussed later, the inverted shape of the magnetic dipole response is responsible for the partial or complete cancellation of an electric mode in reflectivity when both excitations occur at the same frequency.

If the background dielectric function is not too large, it is possible for both $\varepsilon$ and $\mu$ to become simultaneously negative in the frequency spectra. This is the condition for NIR which causes materials to become 'left handed'.[21] The NIR phenomenon has been observed experimentally.[55] The consequence of this condition to the direction of wave propagation and the direction of energy flow can be analyzed qualitatively using $4 \times 4$ matrix formalism. In Case 1, the vector components for the $\vec{k}$ vector [see Eq. (8)] and the Poynting vector, $\vec{S}$ [see Eq. (14) and Eq. (18)] were derived. For each of these vectors, both the $x$ and $z$ components are positive indicating that the wave direction and the direction of energy flow are downward and to the right in the medium (recall that the positive $z$ axis is downward). However, both of these equations change when $\varepsilon \rightarrow -\varepsilon$ and $\mu \rightarrow -\mu$. For the wave vectors, while the $x$ component remains positive, the $z$ component becomes negative. This indicates that the direction of the wave fronts is upward and to the right in the material. For the Poynting vectors, the $x$ component becomes negative while the $z$ component remains positive. This indicates the direction of energy flow is downward to the left in the material. The opposite directions for $\vec{k}$ and $\vec{S}$ as well as their propagation in the third quadrant of the material is now a common understanding for NIR [25]. The qualitative results using $4 \times 4$ matrix formalism further suggest that under conditions of NIR, for crystals with orthorhombic symmetry, the $\vec{k}$ and $\vec{S}$ vectors should diverge as they propagate in the medium in the left handed configuration. Fig. 4(d) simulates this configuration for a material with the negative value of the response functions to those of Section 2 A: $\varepsilon = \left(-4, -6, -8\right)$ and $\mu = \left(-1, -2, -3\right)$.

The implications of the dispersion relations to the magneto-electric case studied in Section II C will now be explored. As explained in that section, a complete description of the $p$ polarized eigenvector required multiplication by $\exp\left(iq_1 z\right)$. This expression can be rewritten in terms of $k_x$, the $x$ component of the incident wave vector:

$$\exp\left(iq_1 z\right) = \exp\left(\frac{i\omega}{2c}\left(\rho + \rho'\right)z\right) \cdot \exp\left(\frac{i\omega}{2c}\left(\sqrt{\left(\rho - \rho'\right)^2 + 4\varepsilon_{xx}\mu_{xx} - \frac{c^2}{\omega^2}\frac{4\varepsilon_{xx}k_x^2}{\varepsilon_{zz}}}\right)z\right).$$ The square root

term can be recognized as $q_a$ as defined in Case 3. Consider the case where $\varepsilon$ and $\mu$ are real and $\rho$ and $\rho'$ are modeled as chiral complex conjugates. The sign of $q_a^2$ will determine the nature of wave propagation in the magneto-electric crystal. For $q_a^2 > 0$, the wave will propagate into the material with sinusoidal amplitude; for $q_a^2 < 0$ the wave will decay exponentially and form an evanescent solution. Following a similar analysis to that for indefinite media outlined in Ref. [25], there will be a value for $k_x$ that makes $q_a = 0$ which is denoted as $k_c$, the cut-off wave vector. This cut off wave vector, which separates propagating waves from decaying waves, can be calculated as: $k_c = \frac{\omega}{2c}\sqrt{\frac{\varepsilon_{zz}}{\varepsilon_{xx}}}\sqrt{(\rho - \rho')^2 + 4\varepsilon_{xx}\mu_{xx}}$. Since anisotropic dispersion relations permit the combinations of $\varepsilon$ and/or $\mu$ to have differing signs, various cases for propagation need to be examined. For example, if $\varepsilon_{xx}\mu_{xx} > 0$ and $\varepsilon_{xx}/\varepsilon_{zz} > 0$, then propagation will occur only if $k_x < k_c$. On the other had, if $\varepsilon_{xx}\mu_{xx} > 0$ and $\varepsilon_{xx}/\varepsilon_{zz} < 0$, there will always be propagation. For $s$ polarized radiation, the cut off conditions will be identical to those in [25] upon adjustment for uniaxial symmetry.

The cut-off analysis can assist in the derivation of the condition for NIR in the magneto-electric crystal examined in Case 3. Recent studies have suggested that this type of magnetic ordering may result in NIR.[10] We consider the case where the damping constants for the response functions are sufficiently small such that all four responses become negative in the same frequency range. From Eq. (28), it can be seen that if $\varepsilon \to -\varepsilon$, $\mu \to -\mu$, $\rho \to -\rho$ and $\rho' \to -\rho'$, the $z$ component of the wave vector will become negative if $k_x < k_c$. From Eq. (32), it can be seen that the same condition causes the sign of the $x$ component of the Poynting vector to become negative. These changes cause the wave vector to propagate upward and to the right in the material while the direction of energy flow will be downward and to the left. Accordingly, when $k_x < k_c$, we expect that NIR can be produced in the magneto-electric crystal.

As explained in Eq. (2), $\rho$ and $\rho'$ may have contributions from both magneto-electric and chiral effects. As explained by Cano[10], the magneto-electric tensor takes on a chiral character for Case 3. In order to illustrate the influence of the chirality on NIR, a material with diagonal tensors $\varepsilon = (-4, -4, -5)$, $\mu = (-3, -3, -4)$, and $\rho_{xy} = i \cdot \xi_{xy} = 3i$ is examined. For these inputs, $q_a^2 = 10.4$ indicating that the wave should propagate without decay in the material. As simulated in Fig. 5(a), the magneto-electric material displays NIR. However, if $\rho$ is changed by only one unit to $\rho = 4$, $q_a^2 = -17.6$ and the cut-off condition is met. As illustrated in Fig. 5(b), the $p$ polarized wave no longer enters the medium while the $s$ polarized wave remains unaffected by the magneto-electric effect. In summary, for proper study of NIR in chiral magneto-electric materials, the interaction of all four response functions must be examined.

The effects of chirality can also be examined for the isotropic symmetry in Case 4. Fig. 5(c) shows the expected results for the wave vectors and Poynting vectors for isotropic $\hat{\varepsilon}$ and $\hat{\mu}$ tensors. When a sufficiently large chiral parameter is introduced, the $s$ polarized wave demonstrates NIR while the $p$ polarized wave remain propagating downward and to the right.

## V. HYBRID MODES AND ADJUSTED OSCILLATOR STRENGTH MATCHING (AOSM)

A rare occurrence of coincident electric and magnetic resonances is possible in, for example, magneto-electric materials, where ligand-field excitations occur due to electronic transitions between $f$-electrons in $RE$ ions.[56] We recently observed this effect in $Dy_3Fe_5O_{12}$ single crystals and have explained it using the concept of the adjusted oscillator strength matching (AOSM) condition.[47] The condition for the matching has been derived for the case of isotropic $\hat{\varepsilon}$ and $\hat{\mu}$ tensors at AOI=0. Below, we expand the theoretical treatment of the AOSM effect for AOI $\neq 0$ and for the case of anisotropic $\hat{\varepsilon}$ and $\hat{\mu}$ tensors.

As discussed in the previous section, the Lorentzian profiles of magnetic and electric dipole excitations have opposing shapes in the reflectivity spectra. A hybrid mode is produced if these modes appear at the same frequency $\omega_h$. We do not consider the magneto-electric effect in the analysis of hybrid modes: $\rho = \rho' = 0$. For hybrid modes, there is an interesting possibility for a partial or complete cancelation of the excitation in the reflectivity spectra $R_s(\omega)$. This motivates the analysis of the derivative $\dfrac{\partial R_s(\omega_h)}{\partial \omega}$ for each mode. Conceptually, for electric and magnetic modes with the same damping coefficient, if their derivatives are identical in magnitude but of opposite sign, then cancellation should occur.

### A. Semi-infinite Configuration and Hybrid Modes

From Eq. (54), the following partial derivative expansion is used for $\dfrac{\partial R_s(\omega_h)}{\partial \omega}$:

$$\frac{dR_{ss}}{d\omega} = r_{ss}^* \left( \frac{\partial r_{ss}}{\partial \varepsilon} \frac{\partial \varepsilon}{d\omega} + \frac{\partial r_{ss}}{\partial \mu} \frac{\partial \mu}{d\omega} \right) + r_{ss} \left( \frac{\partial r_{ss}}{\partial \varepsilon} \frac{\partial \varepsilon}{d\omega} + \frac{\partial r_{ss}}{\partial \mu} \frac{\partial \mu}{d\omega} \right)^* \approx 0 \qquad (56)$$

$r_{ss}$ is the complex reflection coefficient and the Lorentzian oscillator models found in Eq. (53) are used for the response functions, $\varepsilon$ and $\mu$. The same $\gamma_h$ is used for both response functions. When these expressions are inserted into Eq. (56), the following exact derivative can be calculated:

$$\frac{dR_{ss}}{d\omega}\bigg| = r_{ss}\left(\omega_h\right)^* \left(\frac{-1}{\sqrt{\varepsilon\left(\omega_h\right)\mu\left(\omega_h\right)}}\right)\left(\frac{2\omega_h}{\gamma_h}+i\gamma_h\right)\left(\alpha_e^{SI}\left(\omega_h\right)\right)\left[\mu\left(\omega_h\right)S_e + \varepsilon\left(\omega_h\right)S_m \frac{\alpha_m^{SI}\left(\omega_h\right)}{\alpha_e^{SI}\left(\omega_h\right)}\right]$$

$$+ r_{ss}\left(\omega_h\right)\left(\left(\frac{-1}{\sqrt{\varepsilon\left(\omega_h\right)\mu\left(\omega_h\right)}}\right)\left(\frac{2\omega_h}{\gamma_h}+i\gamma_h\right)\left(\alpha_e^{SI}\left(\omega_h\right)\right)\left[\mu\left(\omega_h\right)S_e + \varepsilon\left(\omega_h\right)S_m \frac{\alpha_m^{SI}\left(\omega_h\right)}{\alpha_e^{SI}\left(\omega_h\right)}\right]\right)^*$$

(57)

In Eq. (57), the $\alpha_{e,m}^{SI}$ terms are part of the expressions for $\frac{dr_{ss}}{d\varepsilon}$ and $\frac{dr_{ss}}{d\mu}$, respectively. The superscript SI refers to the semi-infinite configuration. In Eq. (57), we define the bracketed terms, $\mu\left(\omega_h\right)S_e + \varepsilon\left(\omega_h\right)S_m \frac{\alpha_m^{SI}\left(\omega_h\right)}{\alpha_e^{SI}\left(\omega_h\right)}$, as the Adjusted Oscillator Strength (AOS) for reflection, $S_R$. At normal incidence, $\alpha_e^{SI}\left(\omega_h\right) = -\alpha_m^{SI}\left(\omega_h\right)$ and $S_R$ is:

$$S_R\left(\omega_h\right) = \left(\mu\left(\omega_h\right)S_e - \varepsilon\left(\omega_h\right)S_m\right). \tag{58}$$

Eq. (58) suggests that the two modes should cancel in reflection. The condition for complete cancellation is:[47]

$$S_m \varepsilon\left(\omega_h\right) = S_e \mu\left(\omega_h\right) \tag{59}$$

We define Eq. (59) as the Adjusted Oscillator Strength Matching (AOSM) condition. More detail about AOSM and its application to the optical spectra of $Dy_3Fe_5O_{12}$ are available in [47]. When the AOSM condition is satisfied, the electric and magnetic modes interact in such a way as to have no net impact on the background Reflectivity at that point in the spectrum. In other words, the Reflectivity spectra should appear essentially featureless at $\omega_h$. The AOSM condition is also applicable to other dispersion models including, for example, the Coupled Harmonic

Oscillator (CHO) model. We further note that the AOSM condition has similarities to the phenomenon of impedance matching in metamaterials.[57]

Using the complex reflection coefficient, the AOSM condition at a variable AOI can be derived. Using a similar expansion procedure to that above, we get:

$$\frac{\partial r_{ss}}{\partial \omega} \cong \frac{2\omega_h}{\gamma_h^2} \frac{\cos(\theta_0)}{\mu(\omega_h)\sqrt{\varepsilon(\omega_h) - \frac{\sin^2(\theta_0)}{\mu(\omega_h)}}\left(\cos(\theta_0) + \sqrt{\varepsilon(\omega_h) - \frac{\sin^2(\theta_0)}{\mu(\omega_h)}}\right)^2} \left\{ \mu(\omega_h)S_e - \left(\varepsilon(\omega_h) - \frac{2\sin^2(\theta_0)}{\mu(\omega_h)}\right)S_m \right\}.$$

$$(60)$$

From Eq. (60), it can easily be seen that the AOSM condition for variable AOI will be:

$$\mu(\omega_h)S_e = \left(\varepsilon(\omega_h) - \frac{2\sin^2(\theta_0)}{\mu(\omega_h)}\right)S_m \qquad (61)$$

At normal incidence this expression reduces to the formula in Eq. (59), as expected. Eq. (61) also suggests that the AOSM condition may not be satisfied at oblique angles if $\varepsilon(\omega_h)$ is not sufficiently large. At AOI where the AOSM condition is not met, the above equations provide expressions for AOS in reflection which will also assist in the proper characterization throughout the AOI domain.

Also using the complex reflection coefficient, the AOSM condition at normal incidence (AOI=0) can be expanded for an anisotropic material. Since the tensor components which enter into $r_{ss}$ are $\mu_{xx}$ and $\varepsilon_{yy}$, the AOSM condition becomes:

$$S_{m_{xx}}\varepsilon_{yy}(\omega_h) = S_{e_{yy}}\mu_{xx}(\omega_h) \qquad (62)$$

The foregoing analysis relating the AOSM condition and the tendency toward cancellation of modes in Reflectivity can be also qualitatively understood based on Veselago's approach for light propagation in an isotropic, semi-infinite medium with $\mu(\omega) \neq 1$. This approach was discussed previously and involves a simple replacement of the refractive index: for Fresnel's reflection coefficient, $n(\omega) \to \sqrt{\varepsilon(\omega)/\mu(\omega)}$.[21, 22] Using the Lorentzian formulas in Eq. (53), it can be shown that the hybrid resonance is described with an AOS in reflection of $S_R = \left(\mu_\infty \cdot S_e - \varepsilon_\infty \cdot S_m\right)/\mu_\infty^2$.[47]

## B. Thin Film Configuration and Hybrid Modes

For the case of coincident natural frequencies for the magnetic and dielectric oscillators in thin films, the partial derivative expansions for reflectivity and transmission are:

$$\frac{dR_{ss}}{d\omega} \cong r_{ss}^{*}\left(S_2\right) + r_{ss}\left(S_2\right)^{*}$$

$$\frac{dT_{ss}}{d\omega} \cong t_{ss}^{*}\left(S_3\right) + t_{ss}\left(S_3\right)^{*}$$

(63)

For materials with non-negligible film thickness, $S_2$ and $S_3$ are given by:

$$S_2 \approx -\frac{2\omega_h}{\gamma_h^2}\frac{\alpha_e^{R_{TF}}(\omega_h)}{\sqrt{\mu(\omega_h)\varepsilon(\omega_h)}}\left(\mu(\omega_h)S_e - \varepsilon(\omega_h)S_m\right)$$

$$S_3 \approx -\frac{2\omega_h}{\gamma_h^2}\frac{\alpha_e^{T}(\omega_h)}{\sqrt{\mu(\omega_h)\varepsilon(\omega_h)}}\left(\mu(\omega_h)S_e + \varepsilon(\omega_h)S_m\right)$$

(64)

In Eq. (64), the bracketed terms are the Adjusted Oscillator Strengths for thin film reflection and transmission, $S_R$ and $S_T$. Accordingly, the thin film case generates a similar expression for the AOSM condition for reflection. For transmission, on the other hand, the adjusted oscillator strengths are additive. This can also be understood qualitatively based on Veselago's ideas who suggested that if light propagation in transmission is mainly driven by exponential decay and the extinction coefficient, $T_{ss}(\omega)$ becomes a function of $\varepsilon(\omega) \cdot \mu(\omega)$. Using the expansion outlined in Ref. [47], the AOS in transmission is: $S_T \approx S_e \cdot \mu_\infty + S_m \cdot \varepsilon_\infty$ with the two oscillator strengths in $S_T$ being additive. Since most experiments in Transmission are carried out at normal incidence, we do not consider the variable AOI case for the thin film configuration.

The expressions for $S_R$ and $S_T$ allow for analysis of the interesting case of hybrid modes which cancel or disappear in reflectivity but remain strong in transmission combining the magnetic and electric oscillator strengths. The case where hybrid mode magnetic and electric dipole contributions completely cancel in reflection ($S_R = 0$) but add to $S_T$ in transmission requires the solution of the following simultaneous equation:

$$\mu\left(\omega_h\right) S_e - \varepsilon\left(\omega_h\right) S_m = 0$$
$$\mu\left(\omega_h\right) S_e + \varepsilon\left(\omega_h\right) S_m = S_T$$

(65)

Eq. (65) has the approximate solution: $S_e \cong \dfrac{S_T}{2}$ and $S_m \cong \dfrac{S_T}{2\varepsilon_\infty}$. The key implication of Eq. (65) to experimentalists is that experimental data for both Reflectivity and Transmission are needed for proper characterization of a hybrid mode.

# VI. MUELLER MATRIX SIMULATIONS

Based on the foregoing analysis, electric, magnetic, hybrid, electromagnon and chirality excitations in the optical spectra can be simulated  The Mueller Matrices of a chiral multiferroic crystal in 222 point group symmetry (see Case 4) in a semi-infinite configuration are modeled below. The material is assumed to have two main oscillators: a magnetic dipole in $\mu(\omega)$ at 60 cm$^{-1}$ and an electric dipole in $\varepsilon(\omega)$ at 80 cm$^{-1}$. Another hybrid mode that is active in both $\mu(\omega)$ and $\varepsilon(\omega)$ at 70 cm$^{-1}$ is modeled to illustrate the AOSM condition. In addition, a number of scenarios addressing electromagnons and chirality are analyzed in both the frequency and AOI domains. For the diagonal components of the $\hat{\varepsilon}$ and $\hat{\mu}$ tensors, which we consider to be isotropic for this simulation, the Lorentzian models described in Eq. (53) are used. For electromagnon activity, which is also assumed to be isotropic and diagonal, we will cross-examine the following four different dispersion models:

$$\alpha(\omega) = \sqrt{\varepsilon(\omega) \cdot \mu(\omega)} \qquad \alpha'(\omega) = 0 \tag{66}$$

$$\alpha(\omega) = \sqrt{\varepsilon(\omega) \cdot \mu_\infty} \qquad \alpha'(\omega) = 0 \tag{67}$$

$$\alpha(\omega) = \sqrt{\varepsilon_\infty \cdot \mu(\omega)} \qquad \alpha'(\omega) = 0 \tag{68}$$

$$\alpha(\omega) = \alpha'(\omega) = 1 + \sum_{j=1}^{2} \frac{S_{j,a}\omega_{j,a0}{}^2}{(\omega_{j,a0}{}^2 - \omega^2 - i\gamma_{j,a}\omega)} \tag{69}$$

In Eqs. (67) and (68), the magneto-electric resonances occur automatically in $\alpha(\omega)$ at approximately the same frequencies as the resonances in $\varepsilon(\omega)$ and/or $\mu(\omega)$. Note that if $\varepsilon(\omega)$ or $\mu(\omega)$ are Lorentz functions as described in Eq. (53) then $\alpha(\omega)$ in Eq. (67) and Eq. (68) is closely described by the Lorentz function as well. In the model described by Eq. (69), both $\alpha(\omega)$ and $\alpha'(\omega)$ are sums of two oscillators. In our simulations we will describe the chirality effect using a non-resonant linear function of the light frequency

$$\xi(\omega) = -\xi'(\omega) = \omega \cdot X, \qquad (70)$$

where $X$ is the amplitude of chirality. Note that Eq. (70) satisfies the requirement for $\xi(0) \to 0$.

Figure 6 shows the appearance of the two main oscillators in the spectra of the Muller matrix (MM) components at normal incidence (blue spectra). Only the diagonal MM elements are populated due to the absence of the cross polarization terms: $r_{sp} = r_{ps} = 0$. Magnetic and electric dipole excitations at 60 and 80 cm$^{-1}$, respectively, can be easily distinguished in the spectra of the diagonal MM components by their mutually inverted Lorentzian shapes. The oscillator strength of the magnetic dipole at 60 cm$^{-1}$ is more than an order of magnitude less than that for the electrical dipole at 80 cm$^{-1}$, but due to the AOS effect both oscillators appear with similar amplitudes in the diagonal components of the MM. Note that $M_{11}$ represents the spectra of total Reflectivity and the same "enhancement" of the magnetic-dipole features is expected for conventional Reflectance experiments [47]. Figure 6 also illustrates a case of hybrid modes ($h$) with coincident frequency of electric ($e$) and magnetic components at 70 cm$^{-1}$ with $S_e = 0.2$, and $S_m = 0.0168$ (red spectra). The hybrid mode at 70 cm$^{-1}$ practically disappears in the MM spectra due to the AOSM condition described by Eq. (59). With $\varepsilon_\infty = 10$, an initial estimate would suggest that the AOSM condition at 70 cm$^{-1}$ should be satisfied for $S_e = 0.2$ and $S_m = 0.02$. However, it must be remembered, that AOSM occurs with the actual $\varepsilon(\omega_h)$ at the hybrid frequency which is approximately 10.86 in this case. Accordingly, $S_m$ must be less than 0.02 and a nearly perfect matching requires $S_m = 0.0168$.

Figure 7 illustrates the effect of the magneto-dielectric tensor and electromagnon-type excitations on the MM spectra in the frequency domain at AOI=60°. We assume that electromagnon peaks can only appear at the frequencies which correspond to the poles in $\varepsilon(\omega)$ or/and $\mu(\omega)$. We call these excitations elecho-electromagnons (*eem*) and magneto-electromagnons (*mem*), respectively. In addition to the two main oscillators, (*i.e.* magnetic and electric dipoles at 60 and 80 cm$^{-1}$, that were presented in Figure 6), one or two magneto-electric excitations are added using different models for the $\alpha(\omega)$ and $\alpha'(\omega)$ tensors using Eqs. (66), (67), (68), and (69). The magneto-electric resonance that coincides with the magnetic dipole at 60 cm$^{-1}$ is marked *mem*, while the magneto-electric resonance that coincides with the electric dipole at 80 cm$^{-1}$ is marked *eem*. The first three models, which are shown with blue, red, and green spectra have $\alpha'(\omega) = 0$. The blue curve that corresponds to $\alpha(\omega) = \sqrt{\varepsilon(\omega) \cdot \mu(\omega)}$, Eq. (66), has two oscillators at 60 and 80 cm$^{-1}$. As expected, the blue spectra practically overlap with the red ones [$\alpha(\omega) = \sqrt{\varepsilon(\omega) \cdot \mu_\infty}$, Eq. (67)] in proximity of the *eem* resonance at 80 cm$^{-1}$, and also overlap with the green spectra in proximity with the *mem* resonance at 60 cm$^{-1}$.

The diagonal components of MM [blue, red, and green curves in Fig.7(a,b)] are similar to those shown in Fig. 6(a,b) for $\alpha(\omega) = \alpha'(\omega) = 0$. The changes are mostly in the modification of the background values, while the shape of the *mem* and *eem* resonances remains similar to the previous case of no electromagnons [Fig.6(a,b)]. Thus, one can see that the diagonal components of MM which represent conventional Reflectivity spectra are not informative for differentiation between electromagnons and regular magnetic and electric dipoles. In contrast, the off-diagonal elements of the MM in Fig.7(c,d) become populated due to presence of the cross polarization terms: $r_{sp} \neq 0$ and $r_{ps} \neq 0$. Electromagnons revealed themselves as two pronounced peaks at the two resonances in $M_{14}$ and $M_{41}$. These peaks in blue spectra exist because the resonances of $\varepsilon(\omega)$ and $\mu(\omega)$ at 60 and 80 cm$^{-1}$ are automatically incorporated into the equation for $\alpha(\omega) = \sqrt{\varepsilon(\omega) \cdot \mu(\omega)}$. It should be noted that the *mem* and *eem* peaks in $M_{14}$ [Fig. 7(c)] for the models described by Eqs. (67) and (68) have opposite signs: the electromagnon related to the pole in $\mu(\omega)$ at 60 cm$^{-1}$ has a positive amplitude, while the electromagnon related to the pole in $\varepsilon(\omega)$ at 80 cm$^{-1}$ has a negative amplitude. This trend can be useful for identification of the origin of electromagnons as related to electric or magnetic dipoles in optical experiments. It is interesting to cross-evaluate these three models described by Eqs. (66), (67), and (68): the *mem* electromagnon peak is suppressed for $\alpha(\omega) = \sqrt{\varepsilon(\omega) \cdot \mu_\infty}$ (red spectrum), while the *eem* electromagnon peak is enhanced for $\alpha(\omega) = \sqrt{\varepsilon_\infty \cdot \mu(\omega)}$ (green spectrum), if both are compared to the reference $M_{14}$ spectrum for $\alpha(\omega) = \sqrt{\varepsilon(\omega) \cdot \mu(\omega)}$ (blue spectrum). We also analysed the scenario for electromagnons appearing in $\alpha'$ tensor only: $\alpha'(\omega) = \sqrt{\varepsilon(\omega) \cdot \mu(\omega)}$ and $\alpha(\omega) = 0$ and it turns out that this case produces identical spectra of $M_{14}$ to that for $\alpha(\omega) = \sqrt{\varepsilon(\omega) \cdot \mu(\omega)}$ and $\alpha'(\omega) = 0$. In other words, the spectra of the MM components are invariant for the mutual $\alpha(\omega) \Leftrightarrow \alpha'(\omega)$ replacement.

In addition to the three models for electromagnon spectra for $\alpha'(\omega) = 0$, we also considered a SHO model for the case of $\alpha(\omega) = \alpha'(\omega)$, Eq. (69), which is shown with black curves in Figure 7. This model also has two resonances at 60 and 80 cm$^{-1}$ and the corresponding oscillator parameters are $S_{1a} = S_{2a} = 0.09$ and $\gamma_{1a} = \gamma_{2a} = 2$ cm$^{-1}$. The sign of the *eem* resonance in Fig. 7(c,d) has changed compared to the previously considered models with $\alpha'(\omega) = 0$. It is interesting to see that the diagonal componemts of the MM in Fig. 7(a) change quite a lot as well.

For example, the amplitude of the *mem* resonance decreases in $M_{11}$ indicating that there should be another AOSM condition that can result into a complete cancelation of electric and magnetic resonances due to the presence of electromagnons at the same frequencies. The corresponding AOSM conditions for reflectivity spectra in the presence of $\varepsilon(\omega)$, $\mu(\omega)$ and $\alpha(\omega) = \alpha'(\omega) \neq 0$ will be published elsewhere.

For all magneto-dielectric models including $\alpha'(\omega) = 0$ and $\alpha(\omega) = \alpha'(\omega)$, the two off-diagonal MM components $M_{14}$ and $M_{41}$ are of opposite sign. For, $\alpha(\omega) = \alpha'(\omega)$, $M_{14} = -M_{41}$. This relationship is typical for a pure magneto-electric contribution to $\rho$, but it does not hold in the presence of chirality, which in contrast requires $M_{14} = M_{41}$. This observation is important in distinguishing magneto-electric activity from the chiral activity that will be considered below.

Figure 8 illustrates the effect of chiral activity on the MM spectra in the frequency domain at AOI=60°. In addition to the two electric- and magnetic-dipole oscillators (blue spectra), chiral activity is modeled using Eq.(70). Similarly to Fig. 7(c,d), the off-diagonal elements of the MM become populated due to presence of the cross polarization terms. Although the chirality function $\xi(\omega) = \omega \cdot X$ is modeled as a smooth function of $\omega$ without any resonances, the calculated off-diagonal MM components reveal two pronounced Lorentz-shape features marked *mch* and *ech*. These features in $M_{14}$ and $M_{41}$ originate from the main magnetic and electric dipoles at 60 and 80 cm$^{-1}$. It is important to note that *mch* and *ech* peaks in $M_{14}$ and $M_{41}$ [Fig. 8(c,d)] are not inverted with respect to each other as in the case of the electromagnon spectra in Fig. 7(c,d). Thus, the chiral activity in a medium with magnetic and electric dipoles can be observed in the spectra of the off-diagonal MM components. The chiral activity can be distinguished from the magneto-dielectric excitations and electromagnons by measuring the corresponding $M_{14}$ and $M_{41}$ spectra of the MM at oblique AOI. The chiral activity requires $M_{14} = M_{41}$, while the electromagnon-type excitations can be identified if one observes the opposite signs for $M_{14}$ and $M_{41}$. Note, however, that the corresponding reflectivity experiments should be done at an oblique light incidence. As we will see in the following, the back-reflection configuration does not allow differentiation between the chiral and magneto-electric excitations since the chirality effect diminishes at AOI=0 regardless of the magnitude of the parameter $X$ in Eq. (70).

Figure 9 illustrates the MM variation in the AOI domain of the two main electric and magnetic excitations at 80 and 60 cm$^{-1}$ which were previously shown in Fig. 6. The corresponding MM intensities are calculated for 60 cm$^{-1}$ (blue curve) and 80 cm$^{-1}$ (red curve), which correspond to the maximum intensities. The elements $M_{34}$ and $M_{43}$ are non-zero at varying AOI because of the differences between $r_{pp}$ and $r_{ss}$. We note that $M_{34}$ results in opposite signs for each of the two electric-dipole and magnetic dipole resonances, which can help to distinguish between electric and magnetic resonances in the case of, for example, closely-overlaping spectra of magnetic and electric dipoles. Note that the maximum of the difference between magnetic- and electric-dipole signal represented by the red and blue curves in $M_{34} = -M_{43}$ spectra occurs at AOI close to 72.5°, which is the Brewster angle for the simulated scenario in Fig. 9.

Figure 10 illustrates the effect of electromagnons on the MM spectra in the AOI domain. In addition to the two main oscillators, electromagnon excitations are added: *eem* (red curve) and *mem* (blue curve). Since the analysis is done in the AOI domain, it is critical to identify the frequency with which the simulation takes place. We have chosen the two resonance frequencies of 60 cm$^{-1}$ and 80 cm$^{-1}$ for analysis. All off diagonal elements of the MM are populated due to presence of cross polarization terms. Figure 10 suggests that it is possible to distinguish between the *mem* and *eem* excitation through the analysis of $M_{14}$ or $M_{34}$ at the different resonance frequencies. For both $M_{14}$ and $M_{34}$, the *mem* measured at 60 cm$^{-1}$ (blue curve) has the opposite sign to the *eem* measured at 80 cm$^{-1}$ (red curve). As in the case of electro-magnon activity in the frequency domain, we see again that $M_{14}$ and $M_{41}$ have the opposite sign. We further note that even at normal incidence (AOI=0), both $M_{14}$ and $M_{41}$ have non-zero values.

Figure 11 illustrates the effect of chirality on the off-diagonal components of the MM. Variation of the MM intensity at the frequencies of the *mch* and *ech* resonances is shown in the AOI domain. The charality effect is modeled as $\xi(\omega) = X \cdot \omega$, Eq.(70), where $X = 0.015$ cm. In contrast to the no-chirality case presented in Figures 6 and 9, both electric and magnetic dipoles at 60 and 80 cm$^{-1}$ emerged in the off-diagonal MM components. When chirality is active, all 16 elements of the MM are populated. As in the case of chiral activity in the frequency domain, we see again that $M_{14}$ and $M_{41}$ are of the same sign. We also note that, in contrast to Fig 10, both $M_{14}$ and $M_{41}$ are zero at normal incidence, which can be used to distinguish between the pure chiral and pure magneto-electric, or electromagnon, excitations. Accordingly, in both the frequency and AOI domains, it is possible to distinguish between the magneto-electric and chirality effects by examining the relationship between the off-diagonal MM components, *e.g.*, between $M_{14}$ and $M_{41}$.

## VII.  CONCLUSIONS

In this paper, we have used $4 \times 4$ matrix formalism to analyze electromagnetic wave propagation and the optical spectra of complex media. We have demonstrated that their complete description requires the calculation of eigenvalues and eigenvectors of the $\tilde{\Delta}$ matrix using all four response functions. We have used 5 cases to describe the interesting optical effects when the $\varepsilon$, $\mu$ $\rho$, and $\rho'$ tensor components are added to the optical matrix. These effects include birefringence, non-reciprocity, divergence between the wave vector and Poynting vector, NIR, opposing Lorentzian shapes for magnetic and dielectric excitations, hybrid modes and AOSM. For $RE$MnO$_3$ compounds with cycloidal magnetic order (having off diagonal magneto-electric tensors in the dynamic state) analytical expressions have been derived for the eigenvectors, complex reflection coefficients for both the semi-infinite and thin film configurations, and the Poynting vectors. For materials with anisotropic $\hat{\varepsilon}$, $\hat{\mu}$, $\hat{\rho}$ and $\hat{\rho}'$ tensors, the $\tilde{\Delta}$ matrix has been derived. In addition, we have shown how a full Mueller Matrix analysis assists in the proper characterization of the material properties of complex media. For example, although the effects of electro-electromagnons (*eem*) and magneto-electromagnons (*mem*) are difficult to distinguish in the reflectivity spectra, it is possible to distinguish them using full Mueller Matrix analysis over varying AOI. We have also derived the AOSM condition at varying AOI. These derivations will assist in the characterization of metamaterials and multiferroic materials.

## ACKNOWLEDGEMENTS

We gratefully acknowledge the useful discussions with Tae Dong Kang and Tao Zhou at New Jersey Institute of Technology and Adam Dubroka, Prema Marsik, and Christian Bernhard at the University of Fribourg. This work was supported by NSF under Grant No. DMR-0546985.

# FIGURE  CAPTIONS

FIG. 1. Flowchart for the calculation steps in $4 \times 4$ matrix formalism.

FIG. 2. Wave vector diagram for incident and refracted waves propagating in a complex medium.

FIG. 3. (Color online) Wave vector $\vec{k}$ and Poynting vector $\vec{S}$ in for various symmetries and tensor combinations. Unless otherwise indicated, diagonal tensor components are given. $\vec{k}$ for $p$ and $s$ polarizations are solid green and red lines, respectively. $\vec{S}$ for $p$ and $s$ polarizations are dotted green and red lines, respectively.

(a) Case 2: $\hat{\varepsilon} = (4,4,4)$, $\hat{\mu} = (2,2,2)$, $\hat{\rho} = \hat{\rho}' = 0$.

(b) Case 1: $\hat{\varepsilon} = (4,6,8)$, $\hat{\mu} = (1,2,3)$, $\hat{\rho} = \hat{\rho}' = 0$.

(c) Case 3: $\hat{\varepsilon} = (4,4,5)$, $\hat{\mu} = (2,2,3)$, $\rho_{xy} = \rho'_{yx} = 3$.

(d) Case 3: $\hat{\varepsilon} = (4,4,4)$, $\hat{\mu} = (2,2,2)$, $\rho_{xy} = \rho'_{yx} = 3$.

FIG. 4. (Color online) Wave vector $\vec{k}$ and Poynting vector $\vec{S}$ for various symmetries and tensor combinations. Unless otherwise indicated, diagonal tensor components are given in the braskets. $\vec{k}$ for $p$ and $s$ polarizations are solid green and solid red lines, respectively. $\vec{S}$ for $p$ and $s$ polarizations are dotted green and dotted red lines, respectively.

(a) Case 4: $\hat{\varepsilon} = (4,4,4)$, $\hat{\mu} = (2,2,2)$, $\hat{\rho} = \hat{\rho}' = (2.5, 2.5, 2.5)$.

(b) Case 4: $\hat{\varepsilon} = (4,4,4)$, $\hat{\mu} = (2,2,2)$, $\hat{\rho} = \hat{\rho}' = (\sqrt{8}, \sqrt{8}, \sqrt{8})$.

(c) Case 5: $\hat{\varepsilon} = (4,6,8)$, $\hat{\mu} = (3,4,5)$, $\hat{\rho} = \hat{\rho}' = (1,2,3)$.

(d) Case 1: $\hat{\varepsilon} = (-4,-6,-8)$, $\hat{\mu} = (-1,-2,-3)$, $\hat{\rho} = \hat{\rho}' = 0$.

FIG. 5. (Color online) Wave vector $\vec{k}$ and Poynting vector $\vec{S}$ in for various symmetries and tensor combinations. Unless otherwise indicated, diagonal tensor components are given. $\vec{k}$ for $p$ and $s$ polarizations are solid green and solid red lines, respectively. $\vec{S}$ for $p$ and $s$ polarizations are dotted green and dotted red lines, respectively.

(a) Case 3: $\hat{\varepsilon} = (-4, -4, -5)$, $\hat{\mu} = (-3, -3, -4)$, $\rho_{xy} = -\rho'_{yx} = 3i$.

(b) Case 3: $\hat{\varepsilon} = (-4, -4, -5)$, $\hat{\mu} = (-3, -3, -4)$, $\rho_{xy} = -\rho'_{yx} = 4i$.

(c) Case 4: $\hat{\varepsilon} = (2, 2, 2)$, $\hat{\mu} = (1.1, 1.1, 1.1)$, $\hat{\rho} = \hat{\rho}' = 0$.

(d) Case 4: $\hat{\varepsilon} = (2, 2, 2)$, $\hat{\mu} = (1.1, 1.1, 1.1)$, $\hat{\rho} = -\hat{\rho}' = (4i, 4i, 4i)$.

FIG. 6. (Color online) Spectra of MMs for reflectivity at AOI=0 for (a) $M_{11} = M_{22}$, (b) $M_{33} = M_{44}$, (c) $M_{14}$, and (d) $M_{34}$. Only diagonal elements ($M_{11}, M_{22}, M_{33}, M_{44}$) are non-zero in this simulation for isotropic $\varepsilon$ and $\mu$ and $\rho = \rho' = 0$. Blue spectra correspond to the case of two distinct dipole excitations: electric ($e$) at 80 cm$^{-1}$ and magnetic ($m$) at 60 cm$^{-1}$, $\varepsilon_\infty = 10$, $S_e = 0.2$, $S_m = 0.0168$, and $\gamma_e = \gamma_m = 2$ cm$^{-1}$. Note the opposite Lorentzian shapes of the magnetic and electric oscillators in both (a) and (b). Red curves correspond to the same model as that for the blue spectra plus a hybrid mode ($h$) that consists of two additional electric and magnetic excitations with coincident frequency at 70 cm$^{-1}$ with the same parameters: $\varepsilon_\infty = 10$, $S_e = 0.2$, $S_m = 0.0168$, and $\gamma_e = \gamma_m = 2$ cm$^{-1}$. The oscillator strength values for the vanishing hybrid mode ($h$) at 70 cm$^{-1}$ meet the AOSM condition.

FIG. 7. (Color online) Spectra of MMs for reflectivity at AOI=60 $^{\rm o}$ for (a) $M_{11}$, (b) $M_{33}$, (c) $M_{14}$, and (d) $M_{41}$. The off-diagonal elements (*e.g.*, $M_{14}, M_{41}$) are non-zero in this simulation due to the presence of magneto-dielectric tensor. (a) and (b) use left vertical scale, (c) and (d) use right vertical scale. All spectra contain electric and magnetic-dipole excitations at 80 and 60 cm$^{-1}$, $\varepsilon_\infty = 10$, $S_e = 0.2$ and $S_m = 0.0168$ together with two types of electromagnons (*mem* and *eem*) described by different models. Blue spectra: $\alpha(\omega) = \sqrt{\varepsilon(\omega) \cdot \mu(\omega)}$ and $\alpha'(\omega) = 0$, Eq. (66). Red spectra: $\alpha(\omega) = \sqrt{\varepsilon(\omega) \cdot \mu_\infty}$ and $\alpha'(\omega) = 0$, Eq.(67). Green spectra: $\alpha(\omega) = \sqrt{\varepsilon_\infty \cdot \mu(\omega)}$ and $\alpha'(\omega) = 0$, Eq.(68). Black spectra: $\alpha(\omega) = \alpha'(\omega)$, Eq.(69) with two magneto-electric resonances at 60 and 80 cm$^{-1}$, $S_{1a} = S_{2a} = 0.09$ and $\gamma_{1a} = \gamma_{2a} = 2$ cm$^{-1}$. The electromagnons reveal themselves as sharp peaks in $M_{14}$ and $M_{41}$ at 60 and 80 cm$^{-1}$.

FIG. 8.  (Color online) Spectra of MMs for reflectivity at AOI=60 $^{o}$ for (a) $M_{11}$, (b) $M_{33}$, (c) $M_{14}$, and (d) $M_{41}$. The off-diagonal elements (*e.g.*, $M_{14}, M_{41}$) are non-zero in this simulation due to the presence of chirality $\xi(\omega)$. (a) and (b) use left vertical scale, (c) and (d) use right vertical scale. Blue spectra correspond to the reference case of $\xi(\omega) = 0$ and electric and magnetic excitations at 80 and 60 cm$^{-1}$, $\varepsilon_{\infty} = 10$, $S_e = 0.2$ and $S_m = 0.0168$. Two excitations *mch* and *ech* appear in (c) and (d) at the resonant frequencies of 60 and 80 cm$^{-1}$ (red spectra) for $\xi(\omega) = X \cdot \omega$, where $X = 0.015$ cm.

FIG. 9. (Color online) AOI dependence for the MM intensities for the electric excitation at 80 cm$^{-1}$ (red curve) and magnetic excitations and 60 cm$^{-1}$ (blue curve). $\varepsilon_{\infty} = 10$, $S_e = 0.2$ and $S_m = 0.0168$. The off diagonal elements M$_{12}$ and M$_{34}$ become populated at varying AOI because of differences between $r_{pp}$ and $r_{ss}$.

FIG. 10 (Color online) MM of electric and magnetic excitations in the AOI domain at 80 (red curve) and 60 cm$^{-1}$ (blue curve) using $\varepsilon_{\infty} = 10$, $S_e = 0.2$ and $S_m = 0.0168$. In (c) and (d) *mem* and *eem* signal is due to the presence of electromagnons described by $\alpha(\omega) = \sqrt{\varepsilon(\omega) \cdot \mu(\omega)}$ and $\alpha'(\omega) = 0$, Eq. (66).

FIG. 11. (Color online)  MM of electric and magnetic excitations in the AOI domain at 80 (red curve) and 60 cm$^{-1}$ (blue curve) using $\varepsilon_{\infty} = 10$, $S_e = 0.2$ and $S_m = 0.0168$. (a) and (b) use left vertical scale, (c) and (d) use right vertical scale. In (c) and (d) *mch* and *ech* signal is due to the presence of chirality $\xi(\omega) = X \cdot \omega$, Eq.(70), where $X = 0.015$ cm.

# REFERENCES

[1] W. S. Weiglhofer and A. Lakhtakia, *Introduction to Complex Mediums for Optics and Electromagnetics* (SPIE Optical Engineering Press, Bellingham, Wash., 2003).

[2] A. Pimenov, A. A. Mukhin, V. Y. Ivanov, V. D. Travkin, A. M. Balbashov, and A. Loidl, Nat Phys **2**, 97 (2006).

[3] A. B. Sushkov, R. V. Aguilar, S. Park, S. W. Cheong, and H. D. Drew, Phys. Rev. Lett. **98**, 027202 (2007).

[4] R. V. Aguilar, A. B. Sushkov, C. L. Zhang, Y. J. Choi, S. W. Cheong, and H. D. Drew, Physical Review B **76**, 060404 (2007).

[5] Y. Takahashi, Y. Yamasaki, N. Kida, Y. Kaneko, T. Arima, R. Shimano, and Y. Tokura, Physical Review B **79**, 214431 (2009).

[6] J. S. Lee, N. Kida, S. Miyahara, Y. Takahashi, Y. Yamasaki, R. Shimano, N. Furukawa, and Y. Tokura, Physical Review B **79**, 180403 (2009).

[7] A. Shuvaev, F. Mayr, A. Loidl, A. A. Mukhin, and A. Pimenov, ArXiv **1008.2064v1** (2010).

[8] I. V. Lindell, A. H. Sihvola, S. A. Tretyakov, and A. J. Viitanen, *Electromagnetic Waves in Chiral and Bi-Isotropic Media* (Artech House, Boston, 1994).

[9] J. F. Nye, *Physical Properties of Crystals - Their Representation by Tensors and Matrices* (Oxford University Press, Oxford, 2006).

[10] A. Cano, Physical Review B **80**, 180416 (2009).

[11] T. H. O'Dell, *The Electrodynamics of Magneto-Electric Media* (North Holland, Amsterdam, 1970).

[12] J.-P. Rivera, Eur. Phys. J. B **71**, 299 (2009).

[13] E. Georgieva, J. Opt. Soc. Am. A **12**, 2203 (1995).

[14] D. W. Berreman, J. Opt. Soc. Am. **62**, 502 (1972).

[15] R. M. A. Azzam and N. M. Bashara, *Ellipsometry and Polarized Light* (North-Holland, Amsterdam, 1977).

[16] M. Schubert, Thin Solid Films **313-314**, 323 (1998).

17    M. Schubert, *Infrared Ellipsometry on Semiconductor Layer Structures : Phonons, Plasmons, and Polaritons* (Springer, Berlin ; New York, 2004).

18    M. Schubert, T. Hofmann, and C. M. Herzinger, J. Opt. Soc. Am. A **20**, 347 (2003).

19    H. Wöhler, G. Haas, M. Fritsch, and D. A. Mlynski, J. Opt. Soc. Am. A **5**, 1554 (1988).

20    T. G. Mayerhöfer, S. Weber, and J. Popp, Optics Communications **284**, 719 (2010).

21    V. G. Veselago, Sov. Phys. Usp. **10**, 509 (1968).

22    V. Veselago, L. Braginksy, V. Shklover, and C. Hafner, Journal of Computational and Theoretical Nanoscience **3**, 1 (2006).

23    T. M. Grzegorczyk, M. Nikku, X. Chen, B. I. Wu, and J. A. Kong, IEEE Transactions on Microwave Theory and Techniques **53**, 1443 (2005).

24    D. R. Smith, S. Schultz, P. Markoscaron, and C. M. Soukoulis, Physical Review B **65**, 195104 (2002).

25    D. R. Smith and D. Schurig, Phys. Rev. Lett. **90**, 077405 (2003).

26    D. R. Smith, P. Kolinko, and D. Schurig, J. Opt. Soc. Am. B **21**, 1032 (2004).

27    T. Driscoll, G. O. Andreev, D. N. Basov, S. Palit, T. Ren, J. Mock, S.-Y. Cho, N. M. Jokerst, and D. R. Smith, App. Phys. Lett. **90**, 092508 (2007).

28    T. Driscoll, D. N. Basov, W. J. Padilla, J. J. Mock, and D. R. Smith, Phys. Rev. B **75**, 115114 (2007).

29    O. Arteaga, Opt. Lett. **35**, 1359 (2010).

30    O. Arteaga and A. Canillas, Opt. Lett. **35**, 559 (2010).

31    O. Arteaga and A. Canillas, J. Opt. Soc. Am. A **26**, 783 (2009).

32    O. Arteaga and A. Canillas, Opt. Lett. **35**, 3525 (2010).

33    O. Arteaga, E. Garcia-Caurel, and R. Ossikovski, J. Opt. Soc. Am. A **28**, 548.

34    E. Bahar, J. Opt. Soc. Am. B **26**, 364 (2009).

35    E. Bahar, J. Opt. Soc. Am. B **25**, 1294 (2008).

36    E. Bahar, J. Opt. Soc. Am. B **24**, 2807 (2007).

37    E. Bahar, J. Opt. Soc. Am. B **24**, 1610 (2007).

38    E. Bahar, J. Opt. Soc. Am. B **25**, 218 (2008).

39      S. Bassiri, C. H. Papas, and N. Engheta, J. Opt. Soc. Am. A **5**, 1450 (1988).

40      H. Cory and I. Rosenhouse, J. Mod. Opt. **38**, 1229 (1991).

41      E. M. Georgieva, I. J. Lalov, and M. Gospodinov, Optik (Jena) **109**, 173 (1998).

42      A. Konstantinova, K. Konstantinov, B. Nabatov, and E. Evdishchenko, Crystallography Reports **47**, 645 (2002).

43      A. Konstantinova, B. Nabatov, E. Evdishchenko, and K. Konstantinov, Crystallography Reports **47**, 815 (2002).

44      Q. Cheng and T. J. Cui, Physical Review B **73**, 113104 (2006).

45      Q. Cheng and T. J. Cui, J. Opt. Soc. Am. A **23**, 3203 (2006).

46      *http://web.njit.edu/~sirenko/SiMM_web/SiMM.htm. Accessed April, 10, 2011.*

47      P. D. Rogers, et al., Physical Review B **83**, 174407 (2011).

48      P. D. Rogers, T. D. Kang, T. Zhou, M. Kotelyanskii, and A. A. Sirenko, Thin Solid Films **519**, 2668 (2011).

49      E. Hecht, *Optics* (Pearson Education, Inc., 2002).

50      D. Talbayev, A. D. LaForge, S. A. Trugman, N. Hur, A. J. Taylor, R. D. Averitt, and D. N. Basov, Physical Review Letters **101**, 247601 (2008).

51      I. E. Dzyaloshinskii, Sov. Phys. JETP **10**, 628 (1959).

52      W. Eerenstein, N. D. Mathur, and J. F. Scott, Nature **442**, 759 (2006).

53      W. F. Brown, R. M. Hornreich, and S. Shtrikman, Physical Review **168**, 574 (1968).

54      M. Fiebig, T. Lottermoser, D. Frohlich, A. V. Goltsev, and R. V. Pisarev, Nature **419**, 818 (2002).

55      D. R. Smith, W. J. Padilla, D. C. Vier, S. C. Nemat-Nasser, and S. Schultz, Physical Review Letters **84**, 4184 (2000).

56      G. H. Dieke, H. M. Crosswhite, and H. Crosswhite, *Spectra and Energy Levels of Rare Earth Ions in Crystals* (Interscience Publishers, New York,, 1968).

57      A. N. Grigorenko, A. K. Geim, H. F. Gleeson, Y. Zhang, A. A. Firsov, I. Y. Khrushchev, and J. Petrovic, Nature **438**, 335 (2005).

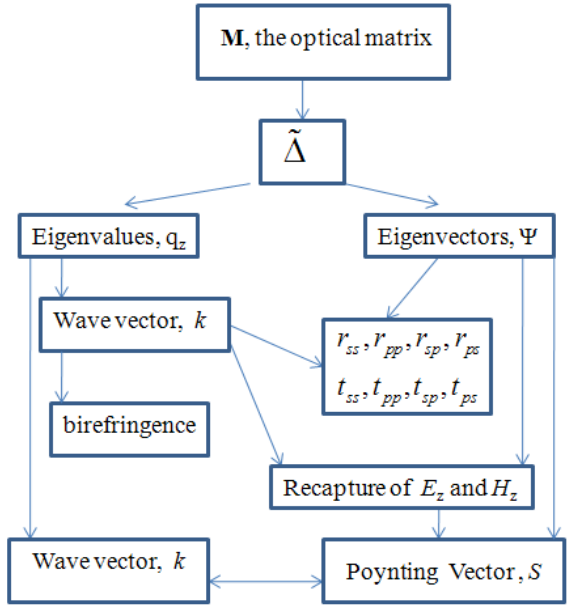

FIG. 1. Flowchart for the calculation steps in $4 \times 4$ matrix formalism.

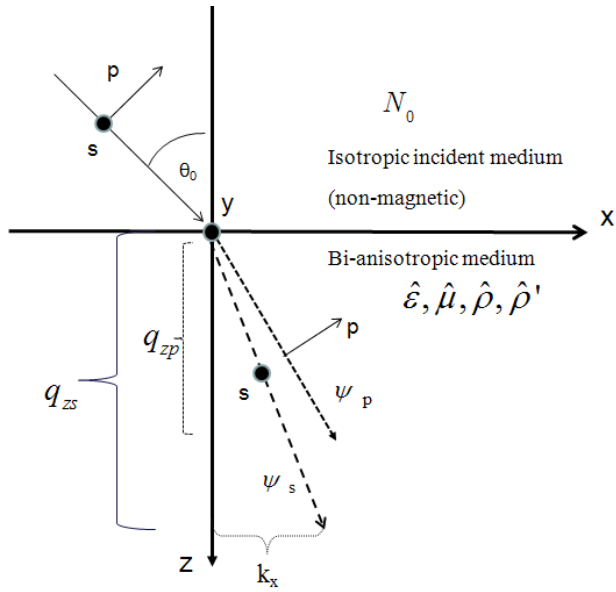

FIG. 2. Wave vector diagram for incident and refracted waves propagating in a complex medium.

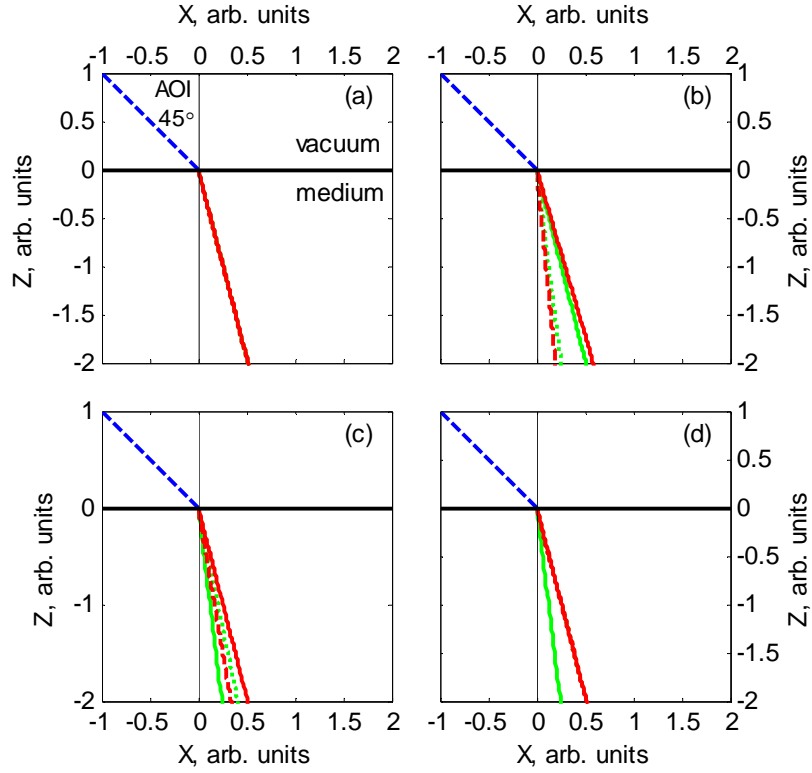

FIG. 3. (Color online) Wave vector $\vec{k}$ and Poynting vector $\vec{S}$ in for various symmetries and tensor combinations. Unless otherwise indicated, diagonal tensor components are given. $\vec{k}$ for $p$ and $s$ polarizations are solid green and red lines, respectively. $\vec{S}$ for $p$ and $s$ polarizations are dotted green and red lines, respectively.

(a) Case 2: $\hat{\varepsilon} = (4,4,4)$, $\hat{\mu} = (2,2,2)$, $\hat{\rho} = \hat{\rho}' = 0$.

(b) Case 1: $\hat{\varepsilon} = (4,6,8)$, $\hat{\mu} = (1,2,3)$, $\hat{\rho} = \hat{\rho}' = 0$.

(c) Case 3: $\hat{\varepsilon} = (4,4,5)$, $\hat{\mu} = (2,2,3)$, $\rho_{xy} = \rho'_{yx} = 3$.

(d) Case 3: $\hat{\varepsilon} = (4,4,4)$, $\hat{\mu} = (2,2,2)$, $\rho_{xy} = \rho'_{yx} = 3$.

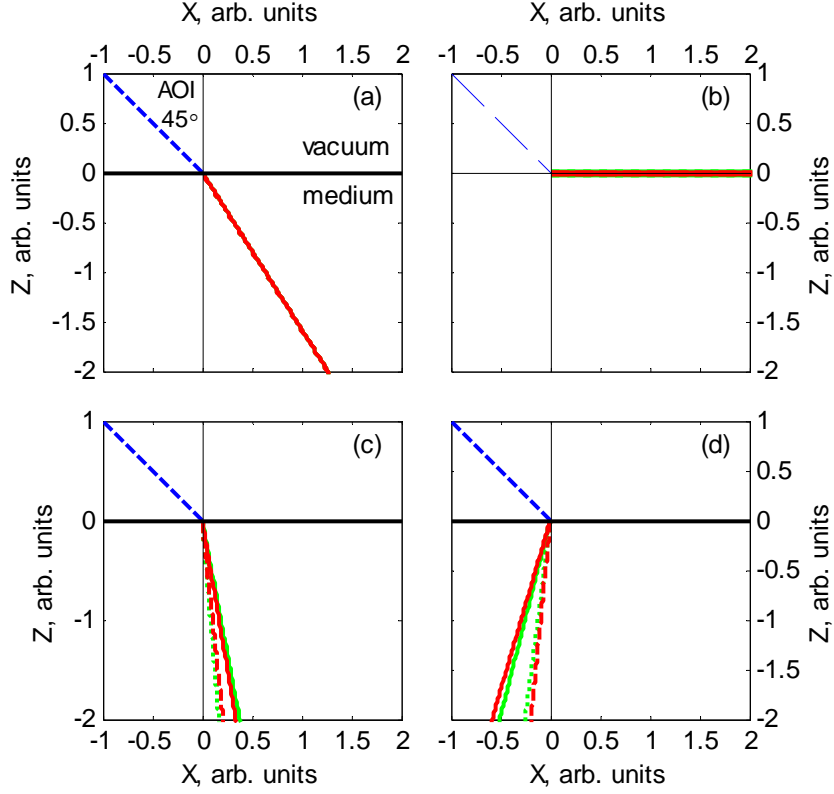

FIG. 4. (Color online) Wave vector $\vec{k}$ and Poynting vector $\vec{S}$ for various symmetries and tensor combinations. Unless otherwise indicated, diagonal tensor components are given in the braskets. $\vec{k}$ for $p$ and $s$ polarizations are solid green and solid red lines, respectively. $\vec{S}$ for $p$ and $s$ polarizations are dotted green and dotted red lines, respectively.

(a) Case 4: $\hat{\varepsilon} = (4,4,4)$, $\hat{\mu} = (2,2,2)$, $\hat{\rho} = \hat{\rho}' = (2.5, 2.5, 2.5)$.

(b) Case 4: $\hat{\varepsilon} = (4,4,4)$, $\hat{\mu} = (2,2,2)$, $\hat{\rho} = \hat{\rho}' = \left(\sqrt{8}, \sqrt{8}, \sqrt{8}\right)$.

(c) Case 5: $\hat{\varepsilon} = (4,6,8)$, $\hat{\mu} = (3,4,5)$, $\hat{\rho} = \hat{\rho}' = (1,2,3)$.

(d) Case 1: $\hat{\varepsilon} = (-4,-6,-8)$, $\hat{\mu} = (-1,-2,-3)$, $\hat{\rho} = \hat{\rho}' = 0$.

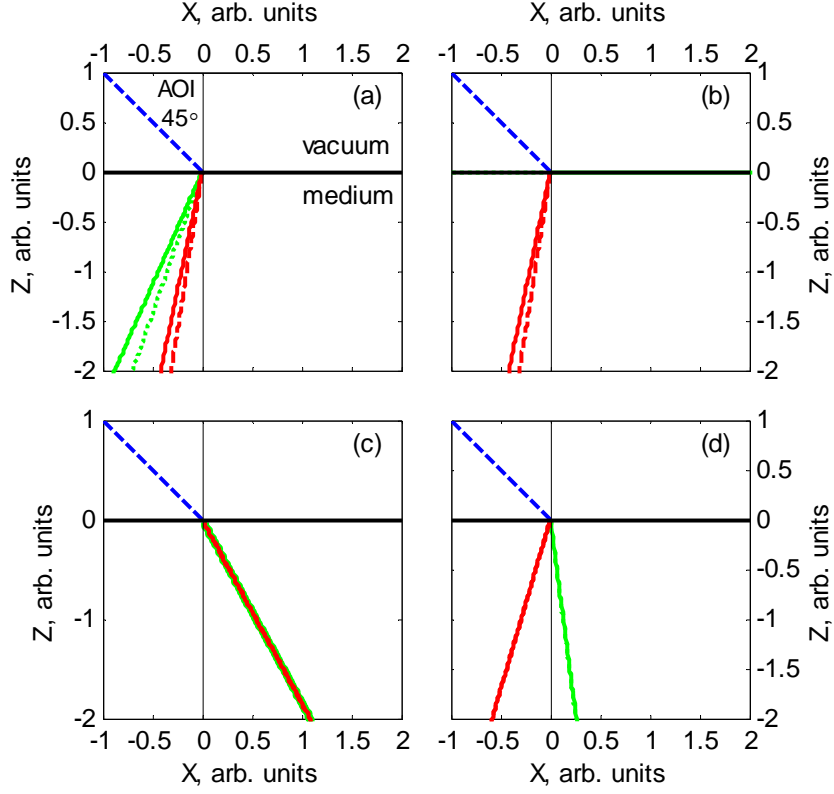

FIG. 5. (Color online) Wave vector $\vec{k}$ and Poynting vector $\vec{S}$ in for various symmetries and tensor combinations. Unless otherwise indicated, diagonal tensor components are given. $\vec{k}$ for $p$ and $s$ polarizations are solid green and solid red lines, respectively. $\vec{S}$ for $p$ and $s$ polarizations are dotted green and dotted red lines, respectively.

(a) Case 3: $\hat{\varepsilon} = (-4,-4,-5)$, $\hat{\mu} = (-3,-3,-4)$, $\rho_{xy} = -\rho'_{yx} = 3i$.

(b) Case 3: $\hat{\varepsilon} = (-4,-4,-5)$, $\hat{\mu} = (-3,-3,-4)$, $\rho_{xy} = -\rho'_{yx} = 4i$.

(c) Case 4: $\hat{\varepsilon} = (2,2,2)$, $\hat{\mu} = (1.1,1.1,1.1)$, $\hat{\rho} = \hat{\rho}' = 0$.

(d) Case 4: $\hat{\varepsilon} = (2,2,2)$, $\hat{\mu} = (1.1,1.1,1.1)$, $\hat{\rho} = -\hat{\rho}' = (4i,4i,4i)$.

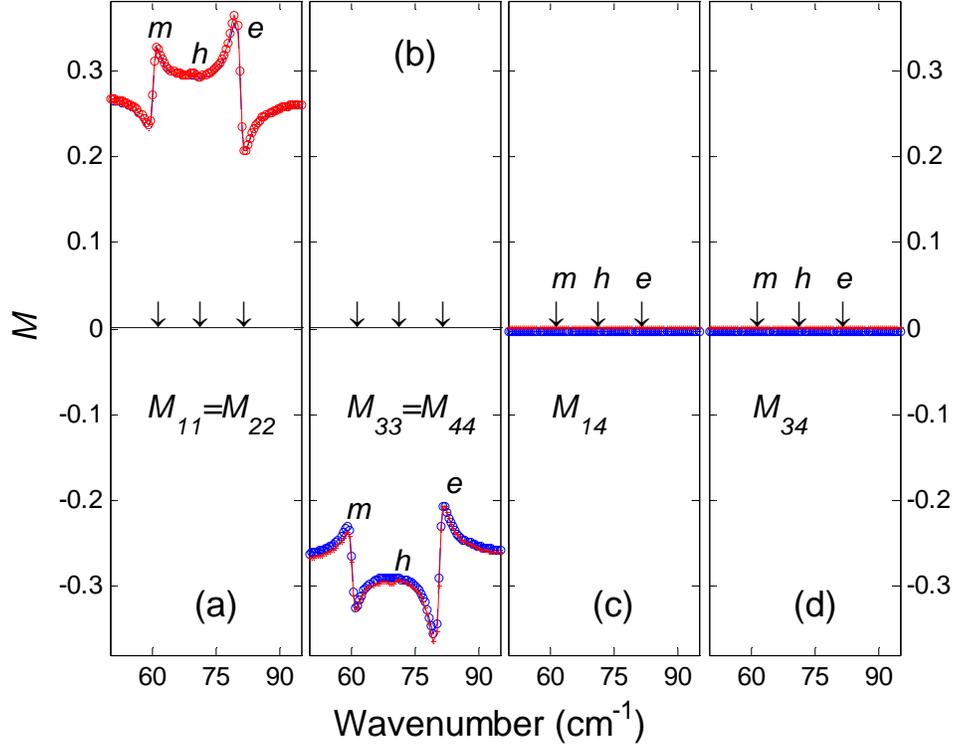

FIG. 6. (Color online) Spectra of MMs for reflectivity at AOI=0 for (a) $M_{11} = M_{22}$, (b) $M_{33} = M_{44}$, (c) $M_{14}$, and (d) $M_{34}$. Only diagonal elements ($M_{11}, M_{22}, M_{33}, M_{44}$) are non-zero in this simulation for isotropic $\varepsilon$ and $\mu$ and $\rho = \rho' = 0$. Blue spectra correspond to the case of two distinct dipole excitations: electric ($e$) at 80 cm$^{-1}$ and magnetic ($m$) at 60 cm$^{-1}$, $\varepsilon_\infty = 10$, $S_e = 0.2$, $S_m = 0.0168$, and $\gamma_e = \gamma_m = 2$ cm$^{-1}$. Note the opposite Lorentzian shapes of the magnetic and electric oscillators in both (a) and (b). Red curves correspond to the same model as that for the blue spectra plus a hybrid mode ($h$) that consists of two additional electric and magnetic excitations with coincident frequency at 70 cm$^{-1}$ with the same parameters: $\varepsilon_\infty = 10$, $S_e = 0.2$, $S_m = 0.0168$, and $\gamma_e = \gamma_m = 2$ cm$^{-1}$. The oscillator strength values for the vanishing hybrid mode ($h$) at 70 cm$^{-1}$ meet the AOSM condition.

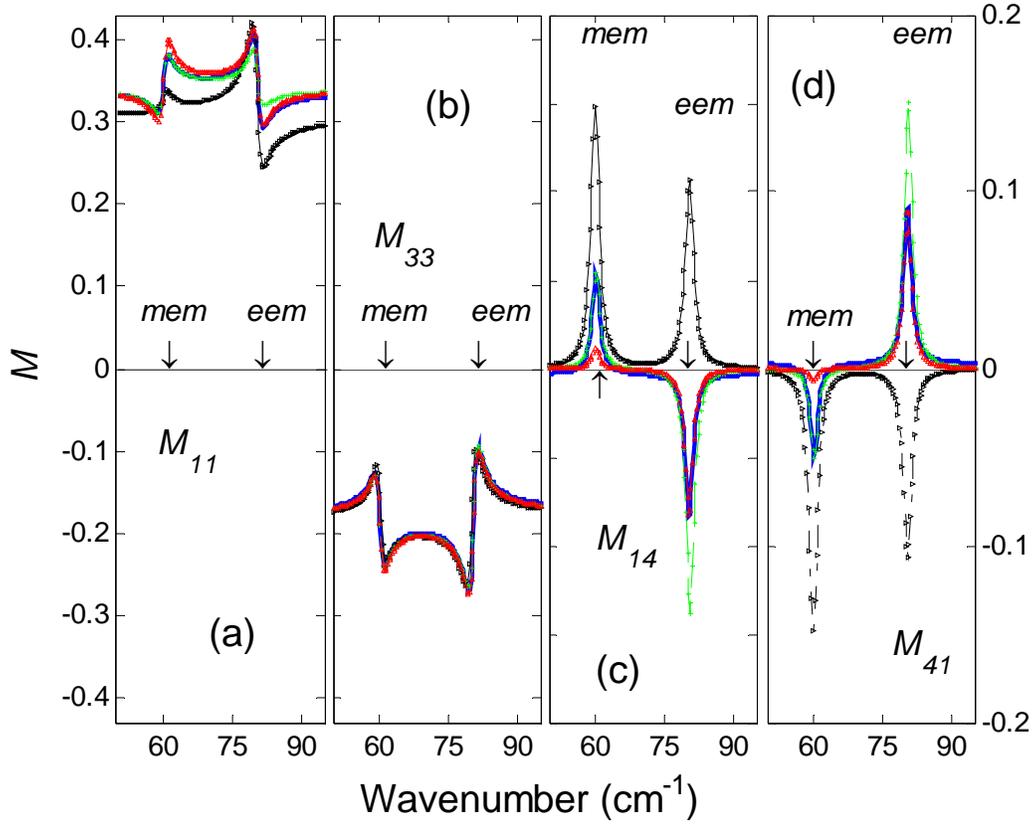

FIG. 7. (Color online) Spectra of MMs for reflectivity at AOI=60° for (a) $M_{11}$, (b) $M_{33}$, (c) $M_{14}$, and (d) $M_{41}$. The off-diagonal elements (*e.g.*, $M_{14}, M_{41}$) are non-zero in this simulation due to the presence of magneto-dielectric tensor. (a) and (b) use left vertical scale, (c) and (d) use right vertical scale. All spectra contain electric and magnetic-dipole excitations at 80 and 60 cm$^{-1}$, $\varepsilon_{\infty} = 10$, $S_e = 0.2$ and $S_m = 0.0168$ together with two types of electromagnons (*mem* and *eem*) described by different models. Blue spectra: $\alpha(\omega) = \sqrt{\varepsilon(\omega) \cdot \mu(\omega)}$ and $\alpha'(\omega) = 0$, Eq. (66). Red spectra: $\alpha(\omega) = \sqrt{\varepsilon(\omega) \cdot \mu_{\infty}}$ and $\alpha'(\omega) = 0$, Eq.(67). Green spectra: $\alpha(\omega) = \sqrt{\varepsilon_{\infty} \cdot \mu(\omega)}$ and $\alpha'(\omega) = 0$, Eq.(68). Black spectra: $\alpha(\omega) = \alpha'(\omega)$, Eq.(69) with two magneto-electric resonances at 60 and 80 cm$^{-1}$, $S_{1a} = S_{2a} = 0.09$ and $\gamma_{1a} = \gamma_{2a} = 2$ cm$^{-1}$. The electromagnons reveal themselves as sharp peaks in $M_{14}$ and $M_{41}$ at 60 and 80 cm$^{-1}$.

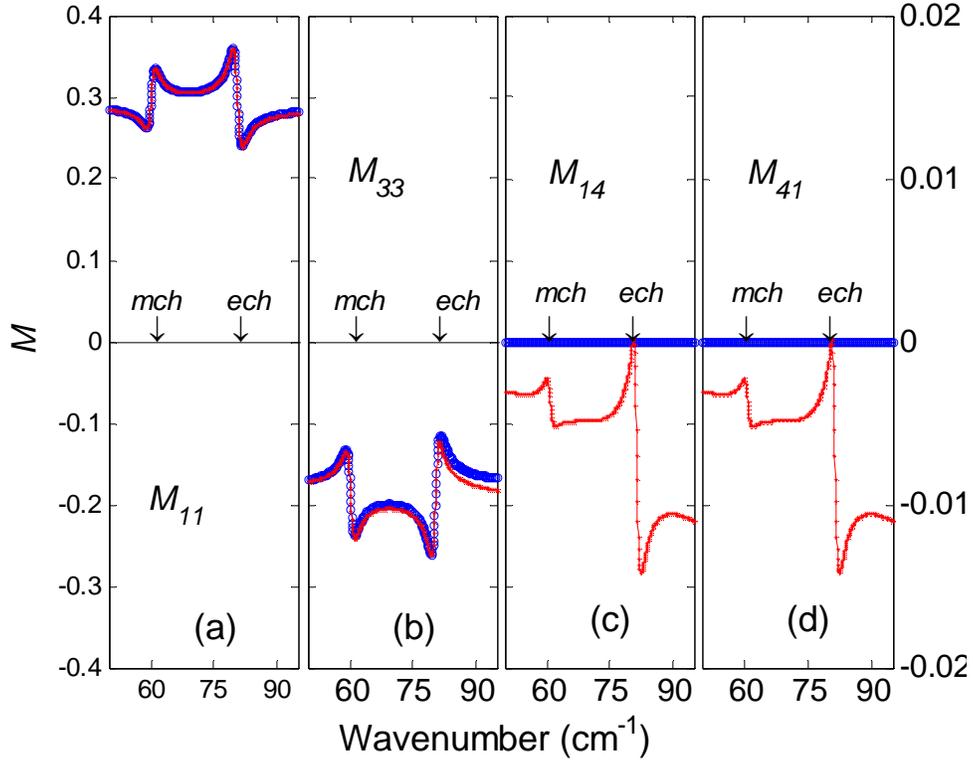

FIG. 8. (Color online) Spectra of MMs for reflectivity at AOI=60 $^{\circ}$ for (a) $M_{11}$, (b) $M_{33}$, (c) $M_{14}$, and (d) $M_{41}$. The off-diagonal elements (*e.g.*, $M_{14}, M_{41}$) are non-zero in this simulation due to the presence of chirality $\xi(\omega)$. (a) and (b) use left vertical scale, (c) and (d) use right vertical scale. Blue spectra correspond to the reference case of $\xi(\omega) = 0$ and electric and magnetic excitations at 80 and 60 cm$^{-1}$, $\varepsilon_{\infty} = 10$, $S_e = 0.2$ and $S_m = 0.0168$. Two excitations *mch* and *ech* appear in (c) and (d) at the resonant frequencies of 60 and 80 cm$^{-1}$ (red spectra) for $\xi(\omega) = X \cdot \omega$, where $X = 0.015$ cm.

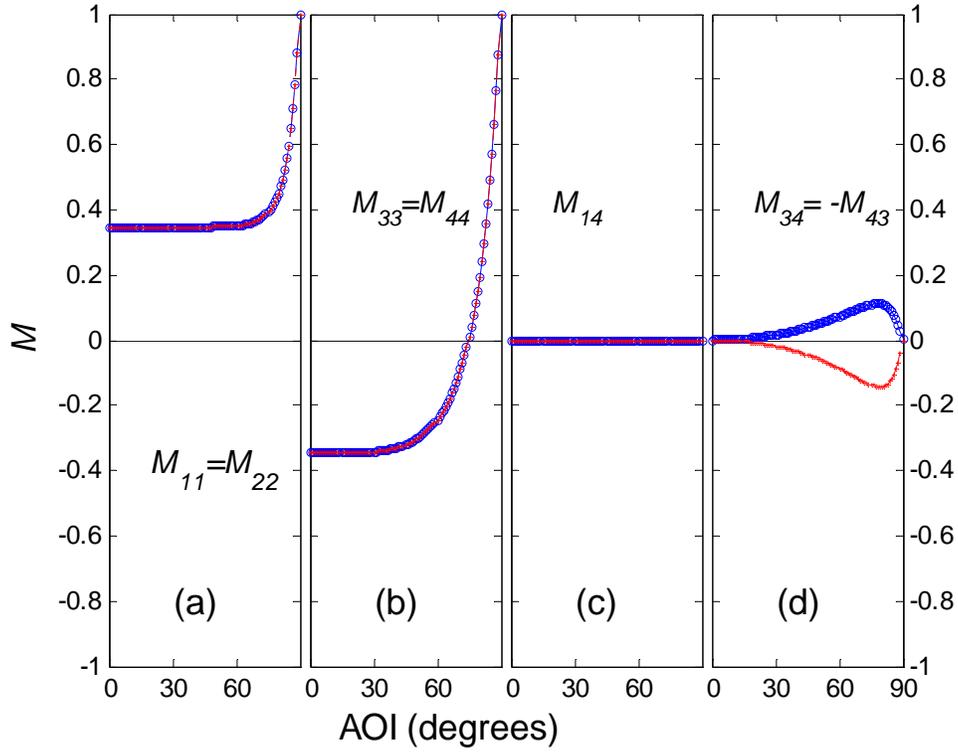

FIG. 9. (Color online) AOI dependence for the MM intensities for the electric excitation at 80 cm$^{-1}$ (red curve) and magnetic excitations and 60 cm$^{-1}$ (blue curve). $\varepsilon_\infty = 10$, $S_e = 0.2$ and $S_m = 0.0168$. The off diagonal elements M$_{12}$ and M$_{34}$ become populated at varying AOI because of differences between $r_{pp}$ and $r_{ss}$.

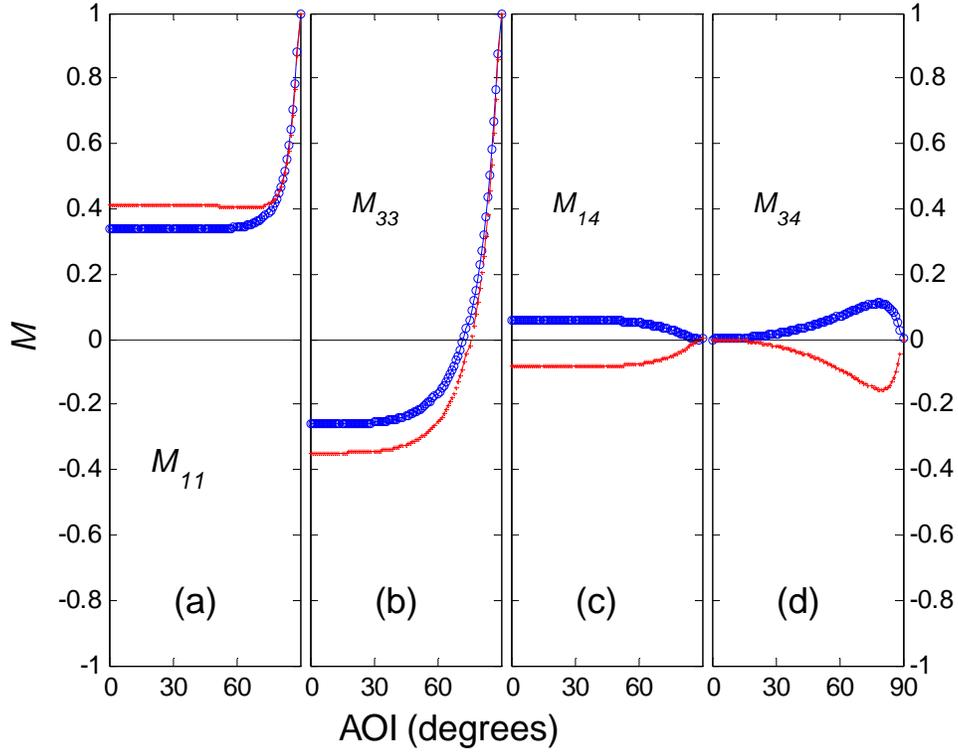

FIG. 10 (Color online) MM of electric and magnetic excitations in the AOI domain at 80 (red curve) and 60 cm$^{-1}$ (blue curve) using $\varepsilon_\infty = 10$, $S_e = 0.2$ and $S_m = 0.0168$. In (c) and (d) *mem* and *eem* signal is due to the presence of electromagnons described by $\alpha(\omega) = \sqrt{\varepsilon(\omega) \cdot \mu(\omega)}$ and $\alpha'(\omega) = 0$, Eq. (66).

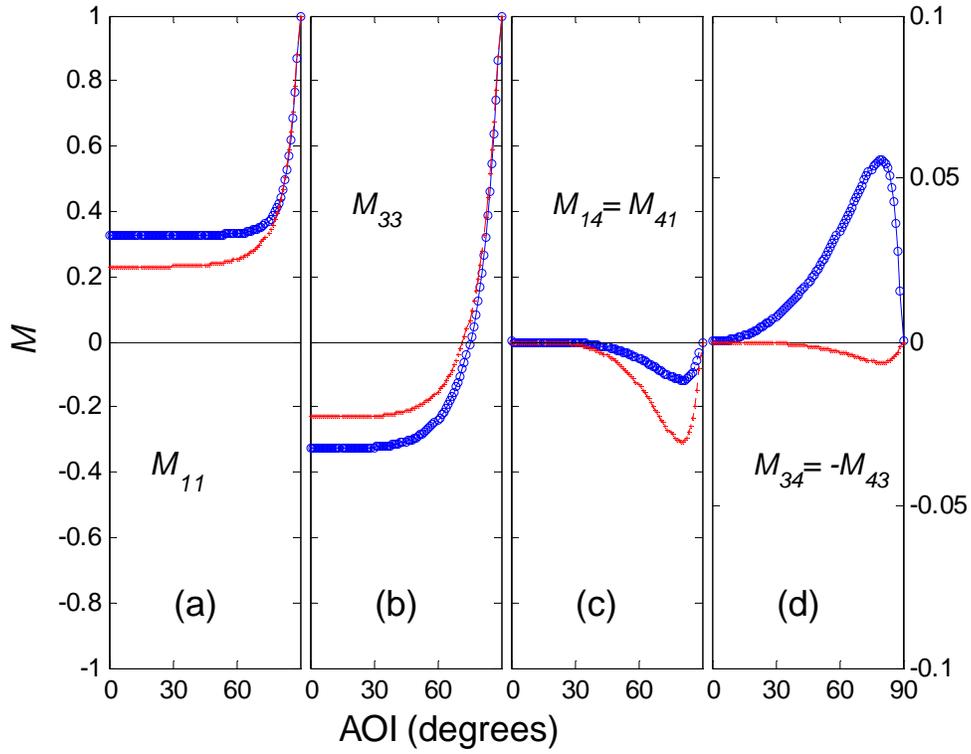

FIG. 11. (Color online) MM of electric and magnetic excitations in the AOI domain at 80 (red curve) and 60 cm$^{-1}$ (blue curve) using $\varepsilon_\infty = 10$, $S_e = 0.2$ and $S_m = 0.0168$. (a) and (b) use left vertical scale, (c) and (d) use right vertical scale. In (c) and (d) *mch* and *ech* signal is due to the presence of chirality $\xi(\omega) = X \cdot \omega$, Eq.(70), where $X = 0.015$ cm.